\documentclass[letterpaper,twocolumn,10pt]{article}
\usepackage{usenix2019ATC,epsfig,endnotes}
\usepackage{graphicx}
\usepackage{epstopdf}
\usepackage{subfig}
\begin{document}

\date{}


\title{\Large \bf SAWL:A Self-adaptive Wear-leveling NVM Scheme for High Performance Storage Systems}

\author{
	{\rm Jianming Huang{\textsuperscript{*}}, Yu Hua{\textsuperscript{*}}, Pengfei Zuo{\textsuperscript{*}}, Wen Zhou{\textsuperscript{*}}, Fangting Huang{\textsuperscript{*}}}
	\\
	{\textsuperscript{*}}Huazhong University of Science and Technology \\
}

\maketitle

\subsection*{Abstract}
In order to meet the needs of high performance computing (HPC) in terms of large memory, high throughput and energy savings, the non-volatile memory (NVM) has been widely studied due to its salient features of high density, near-zero standby power, byte-addressable and non-volatile properties. In HPC systems, the multi-level cell (MLC) technique is used to significantly increase device density and decrease the cost, which however leads to much weaker endurance than the single-level cell (SLC) counterpart. Although wear-leveling techniques can mitigate this weakness in MLC, the improvements upon MLC-based NVM become very limited due to not achieving uniform write distribution before some cells are really worn out. To address this problem, our paper proposes a self-adaptive wear-leveling (SAWL) scheme for MLC-based NVM. The idea behind SAWL is to dynamically tune the wear-leveling granularities and balance the writes across the cells of entire memory, thus achieving suitable tradeoff between the lifetime and cache hit rate. Moreover, to reduce the size of the address-mapping table, SAWL maintains a few recently-accessed mappings in a small on-chip cache. Experimental results demonstrate that SAWL significantly improves the NVM lifetime and the performance for HPC systems, compared with state-of-the-art schemes.

\vspace{-0.3cm}
\section{Introduction}
High performance computing  (HPC) systems generally require large-size memory, high I/O throughput and significant energy savings. Due to meeting all these needs, non-volatile memory (NVM) has been widely used in high performance systems~\cite{markthub2018dragon,kim2017papyruskv,peng2018siena,wu2018runtime}. For HPC applications, fitting the larger workloads in NVM than DRAM can efficiently alleviate the constraints from memory space and reduce the data movements between high-speed memory and low-speed disks to deliver high performance~\cite{markthub2018dragon}. Moreover, the recent measurements of the Intel Optane DC Persistent Memory Module demonstrate the significant performance improvements upon typical real-world applications~\cite{izraelevitz2019basic}. Existing studies ~\cite{lefurgy2003energy,memorypower} have also shown that leakage energy grows with the memory capacity, dissipating as much heat as dynamic energy and becomes a main contributor to operational costs. NVM technologies
\cite{chen2010advances,chen2003access,endoh2016overview,kultursay2013evaluating,lee2010phase}, such as STT-RAM, PCM, and RRAM, hence become promising and important for HPC applications.

In practice, NVM fails to achieve high performance and actually increases the complexity of management due to the limited lifetime, which causes frequent update and re-configurations. The property of limited lifetime has become the performance bottleneck of storage systems, especially for HPC applications that usually contain large amounts of write operation. Moreover, in order to offer large space capacity and relatively cheap costs, device vendors often provide multi-level-cell (MLC)-based NVM for real-world applications. Compared with single-level-cell (SLC)-based NVM, MLC-based NVM exhibits higher storage density, lower costs and comparable read latency, thus achieving better performance in memory-sensitive HPC applications. Unfortunately, the lifetime of MLC becomes exacerbated, since MLC stores more bits in a single cell and results in weak endurance.
The MLC technique used in NVM (PCM and RRAM) is able to support the rapid growth in device capacity and density but at the cost of much weaker endurance than the SLC counterpart. The advanced fabrication technique in MLC packs more than one bit in a single cell \cite{lee2012multi}, thus allowing NVM to achieve ultra-high density. However, due to the iterative program-and-verify (P\&V) technique, the MLC technology produces remarkable variations on access latency and cell endurance. Table \ref{table1} summarizes the characteristic parameters of SLC- and MLC-based NVM technologies. 

Compared with SLC, the MLC-based NVM increases access latency by 2$\sim$4 times and decreases endurance by 100 times due to unavoidable over-programming operation. For example, the SLC PCM devices are expected to last for ${10}^{7}\sim{10}^{8}$ writes per cell \cite{freitas2008storage,zhao2014slc}, and the RRAM technology has a per-cell write limit between ${10}^{8}$ and ${10}^{12}$ in the SLC mode. But the cell endurance of MLC PCM only reaches ${10}^{5}\sim{10}^{6}$ writes per cell~\cite{gleixner2009reliability}, and that of the MLC RRAM decreases to ${10}^{7}$ writes per cell \cite{lee2012multi}.

\begin{table}[h]
\linespread{1.4}
\centering
\scriptsize
\setlength{\abovecaptionskip}{0pt}
\setlength{\belowcaptionskip}{5pt}
\caption{Key features of SLC- and MLC-based NVM technologies.}
\label{table1}
\begin{tabular}{|l|c|c|c|c|}
\hline
              & SLC PCM   & SLC RRAM & MLC PCM    & MLC RRAM \\ \hline
Read latency  & 150ns & 10ns   & 250ns      & 50ns    \\ \hline
Write latency & 450ns     & 50ns & 1.5us & 350ns    \\ \hline
Cell endurance     & $10^{7} \sim 10^{8}$   & $10^{8} \sim 10^{12}$ & $10^{5} \sim 10^{6}$    & $\sim$ $10^{7}$      \\ 
\hline
\end{tabular}
\end{table}

In order to extend the lifetime of MLC-based NVM, the wear-leveling technique attempts to make write operation uniformly distributed by frequently remapping logical lines to new physical positions, which can also prevent brute-force attacks to a certain physical line. {However, we observe that existing wear-leveling algorithms \cite{qureshi2009enhancing,seong2010security,seznec2009towards,zhao2014slc,zhao2014leveling} initially designed for SLC-based NVM, become inefficient in MLC-based NVM systems. Specifically,} to prevent the malicious attacks~\cite{seong2010security} that guess the physical location and continuously wear a given line, existing algorithms perform the remapping in the randomized manner without recording the accurate write counts of memory cells. Hence, they attempt to achieve wear leveling by randomly shuffling logical-physical address mappings via algebraic functions to evenly disperse the logical lines written most frequently to as many physical lines as possible. This requires a huge number of rounds of data exchanges before a probabilistically uniform distribution of write counts of all physical lines in an NVM can be achieved \cite{yu2012increasing,qureshi2011practical}. 
However, in practice, the low endurance of MLC-based NVM implies that some cells can be worn out long before this uniform distribution is achieved. {As a result,} existing work fails to attain long lifetime of MLC-based NVM (the quantitative analysis is shown in Section 2).

There are two straightforward solutions to accelerate data exchanges and avoid some lines being worn out before being swapped. One is to increase the exchange frequency. However, frequent content exchanges increase write amplification and block the data access, which in turn significantly decrease performance and increase energy consumption.

The other is to decrease the wear-leveling granularities (e.g., \emph{region size}) to mitigate the imbalanced writes across the entire memory, which however significantly increases the size of address mapping table (e.g., hundreds of megabytes). Therefore, the mapping table is too large to be fully held into the on-chip cache which leads to severe performance degradation due to the long latency of address translation.

{To address this problem,} a tiered architecture can be considered,  which stores the entire address mapping table in the main memory (DRAM or NVM devices) and holds the recently-accessed entries in a small on-chip SRAM cache. {In fact, this intuitive solution} often fails to provide sufficient performance improvements for the applications with substantial random access patterns due to the low cache hit rate of HPC applications. {Hence,} we propose a self-adaptive wear-leveling scheme (SAWL) that dynamically changes the wear-leveling granularities to accommodate more useful addresses in the cache, thus significantly improving cache hit rate. {As a result,} SAWL is able to achieve both long lifetime and high performance. The main contributions are summarized:

\begin{enumerate}
  \item {\emph{\textbf{Insights for wear-leveling schemes on MLC-based NVM.}} We investigate the effectiveness that state-of-the-art wear-leveling algorithms work on MLC-based NVM, including table-based wear-leveling (TBWL)~\cite{zhou2009durable}, algebraic-based wear-leveling (AWL)~\cite{qureshi2009enhancing,seong2010security}, and hybrid wear-leveling (HWL) schemes~\cite{seznec2009towards,yu2012increasing}. We observe that TBWL and AWL have the vulnerability of either Repeated Address Attack (RAA) or significant NVM lifetime reduction. HWL is able to achieve high lifetime but causes significant on-chip storage overhead to store address mappings.}
  \item {\emph{\textbf{An efficient wear-leveling scheme for MLC-based NVM.}} We propose a Self-Adaptive Wear-Leveling (SAWL) scheme for MLC-based NVM. SAWL maintains recently-accessed address mappings in a small on-chip cache managed by the memory controller. To improve the cache hit rate, SAWL dynamically changes the wear-leveling granularities by means of region-merge and region-split operations as shown in Section 3.2. As a result, SAWL is able to achieve both high lifetime and performance.}
  \item {\emph{\textbf{Implementation and evaluation.}} We have implemented SAWL and evaluated it using the gem5~\cite{binkert2011gem5}. Experimental results show that SAWL improves $25\%\sim51\%$ ($50\%\sim78\%$) of ideal lifetime for the MLC-based NVM system with ${10}^{6}$ (${10}^{5}$) cell endurance, compared with state-of-the-art wear-leveling schemes. Moreover, existing wear-leveling schemes incur 25\% IPC decrease on average, while SAWL only decreases the IPC performance by 5\% on average, compared with a baseline system without any wear-leveling algorithms.}
\end{enumerate}

The rest of the paper is organized as follows. Section 2 introduces the background and motivation. The design of SAWL is described in Section 3. Section 4 presents the evaluation results and analysis. Section 5 presents the related work. We conclude this paper in Section 6.

\vspace{-0.3cm}

\section{Background and Motivation}

In this section, we present the background on wear leveling in NVM to facilitate our discussion and analyze the important observations that motivate our SAWL design.

\subsection{Existing Wear-Leveling Algorithms}

{Wear-leveling schemes are proposed to extend the lifetime of NVM and defend against security attacks by uniformly distributing writes among all NVM cells.} When a region has been written for a certain amount, the wear-leveling algorithm is performed to exchange the data in/beyond this region. The number of the writes to trigger the wear-leveling is called swapping period.
According to the mapping relationship between the logical and physical addresses, existing wear-leveling schemes can be classified into three categories: table-based wear-leveling (TBWL), algebraic-based wear-leveling (AWL), and hybrid wear-leveling (HWL) schemes. Wear-leveling is transparent for upper-level applications due to the mapping relationship between the logical and physical addresses. Applications can simply access the same contents according to the same logical addresses and overlook the physical addresses where data are actually stored.

\begin{figure}[t]
\vspace{-0.3cm}
\centering
\includegraphics[width=0.45\textwidth]{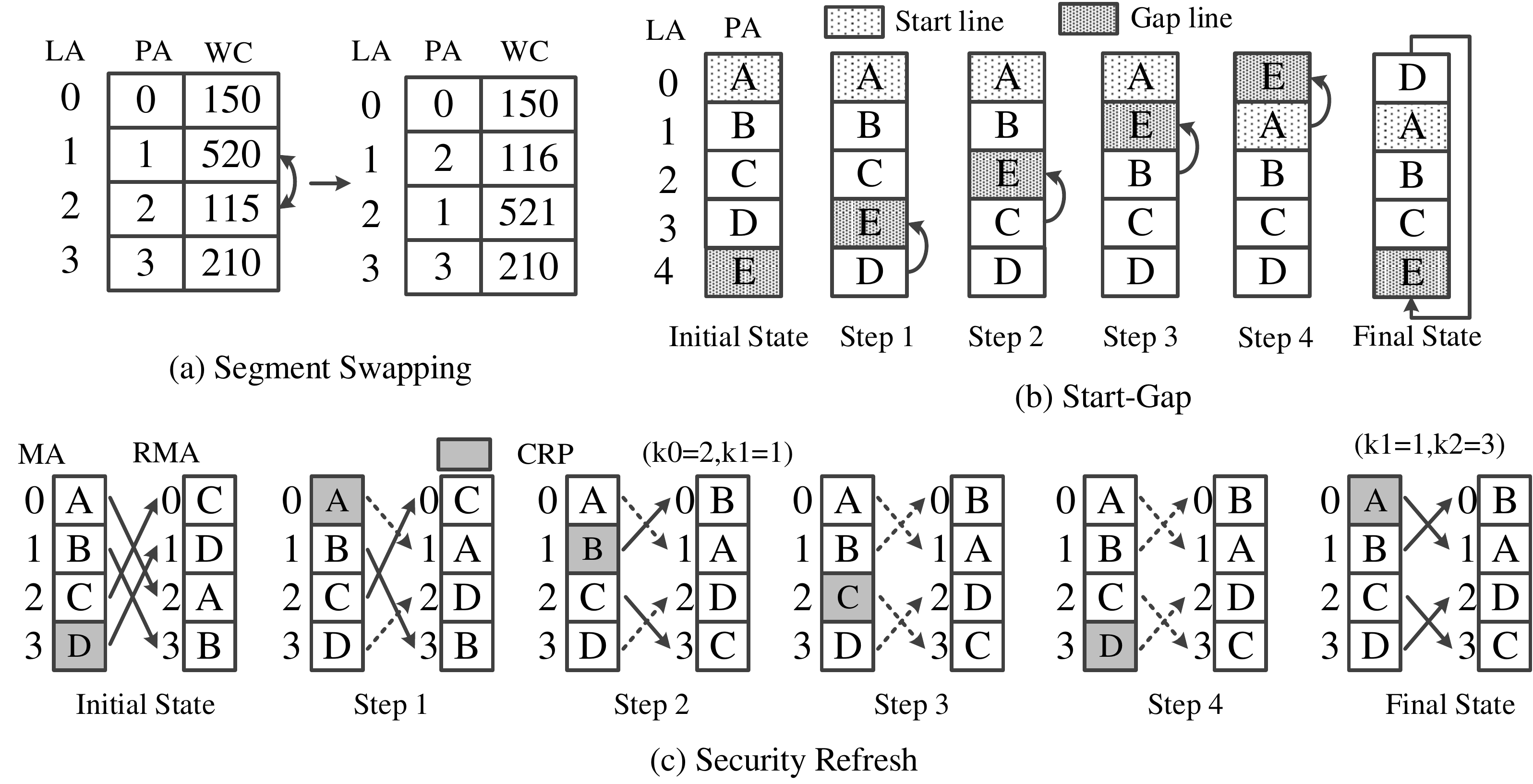}
\caption{Table-based and algebraic wear-leveling schemes ((a) is TBWL scheme, (b) and (c) are AWL schemes).}
\label{table_awl_policies}
\vspace{-0.45cm}
\end{figure}

{\emph{TBWL schemes}, e.g., Segment Swapping~\cite{zhou2009durable}, record} the corresponding mapping relationship between a logical line address (LA) and its physical counterpart (PA). When the write count (WC) of one PA triggers the wear-leveling, Segment Swapping exchanges the data between this PA and the least used PA in the same region, as shown in Fig.~\ref{table_awl_policies}(a). A line is the atomic memory-access unit whose size is equal to that of the last-level cache line. This, however, results in a huge space overhead in keeping track of the mapping information in all memory lines.

{\emph{AWL schemes}} leverage algebraic mapping functions to randomly generate the physical address for a given logical address. The space overhead is extremely low since the algebraic function using space-efficient hardware structure replaces the address-mapping table in the table-based wear-leveling algorithms. The representative AWL schemes include region-based Start-Gap (RBSG)~\cite{qureshi2009enhancing} and two-level Security Refresh (TLSR)~\cite{seong2010security}, as shown in Fig.~\ref{table_awl_policies}(b) and~\ref{table_awl_policies}(c).
{RBSG always swaps a memory line with its neighboring line, which is easily attacked by maliciously-contrived code through simple buffer-overflow detection ~\cite{seong2010security}.}
To defend against such malicious attacks, {TLSR uses} dynamically generated random keys and XOR operations to change address mappings in a more unpredictable way to reduce the security vulnerability of TLSR. 
{However,} as the number of regions increases, a pure AWL scheme usually fails to balance write traffic among the regions, which enables the lines of the heavily-written regions to be worn out much earlier than others.

\begin{figure}[t]
\centering
\includegraphics[width=0.4\textwidth]{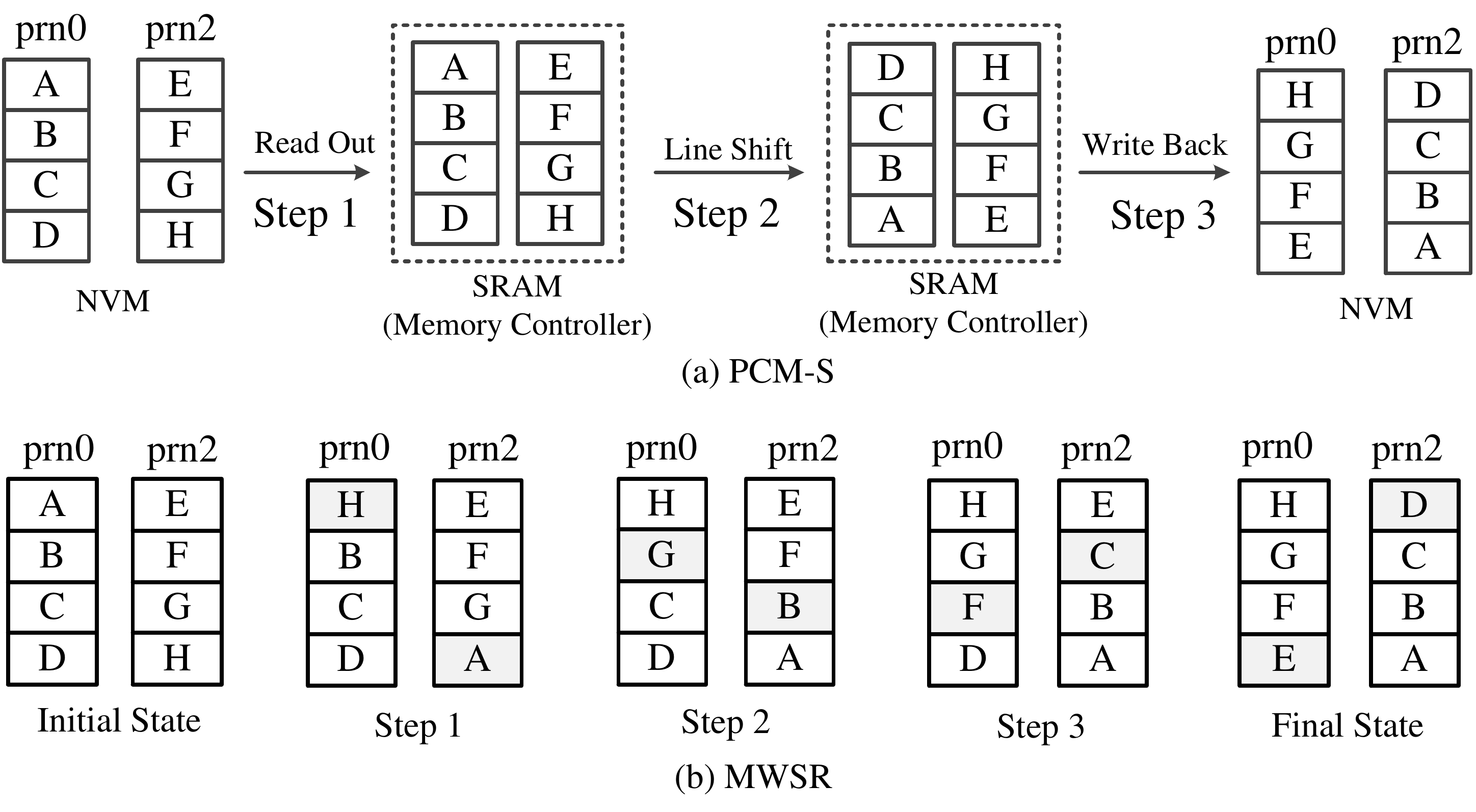}
\caption{The state-of-the-art hybrid wear-leveling schemes.}
\label{hwl_policies}
\vspace{-0.3cm}
\end{figure}

{\emph{HWL schemes}} combine the algebraic and table-based wear-leveling algorithms, such as PCM-S ~\cite{seznec2009towards} and MWSR ~\cite{yu2012increasing} as shown in Fig.~\ref{hwl_policies}, which use a mapping table to keep track of the mapping relationship between the logical region address of a line and the physical region address of its corresponding physical line, and leverage the algebraic function to obtain the physical location of lines within each region according to the given logical address. In general, the physical address offset ($pao$) of the memory lines within the region can be obtained through $pao = lao \bigoplus key$, where $lao$ represents the logical address offset and $key$ denotes the offset parameter within a region. The HWL algorithms disperse writes across the entire memory by randomly exchanging the regions and shifting the location of its lines simultaneously~\cite{seznec2009towards}.

\subsection{{Problems of Wear-leveling Algorithms on MLC-based NVM}}

{The above-mentioned wear-leveling algorithms work well for SLC-based NVM. However, we observe that these algorithms expose strong security vulnerability and shortened lifetime for MLC-based NVM, due to decreased cell endurance and increased device capacity of MLC-based NVM, as elaborated next.}

\begin{itemize}
  \item \textbf{ Decreased cell endurance.} The MLC technique decreases NVM cell endurance by two orders of magnitude ~\cite{gleixner2009reliability,lee2012multi}. This weakened endurance leads to insufficient numbers of data exchanges across the entire memory for the existing wear-leveling algorithms because the number of data exchanges is proportional to the cell endurance, which results in serious write imbalance and severely reduces the lifetime of MLC-based NVM systems.

  \item \textbf{Increased device capacity.} The given trend suggests that the capacity of a single bank and an NVM system is likely to increase potentially exponentially with the advanced manufacturing technology and multithreaded application requirements. Thus, to ensure sufficient data exchanges within and among regions in the entire memory space, the wear-leveling algorithms must increase the number of regions, a number that is proportional to NVM capacity. However, as the number of regions increases, the hardware overhead increases proportionally. The space and hardware overhead can become unacceptably high for the practical systems.
\end{itemize}

\begin{figure}[t]
\centering
\includegraphics[width=0.45\textwidth]{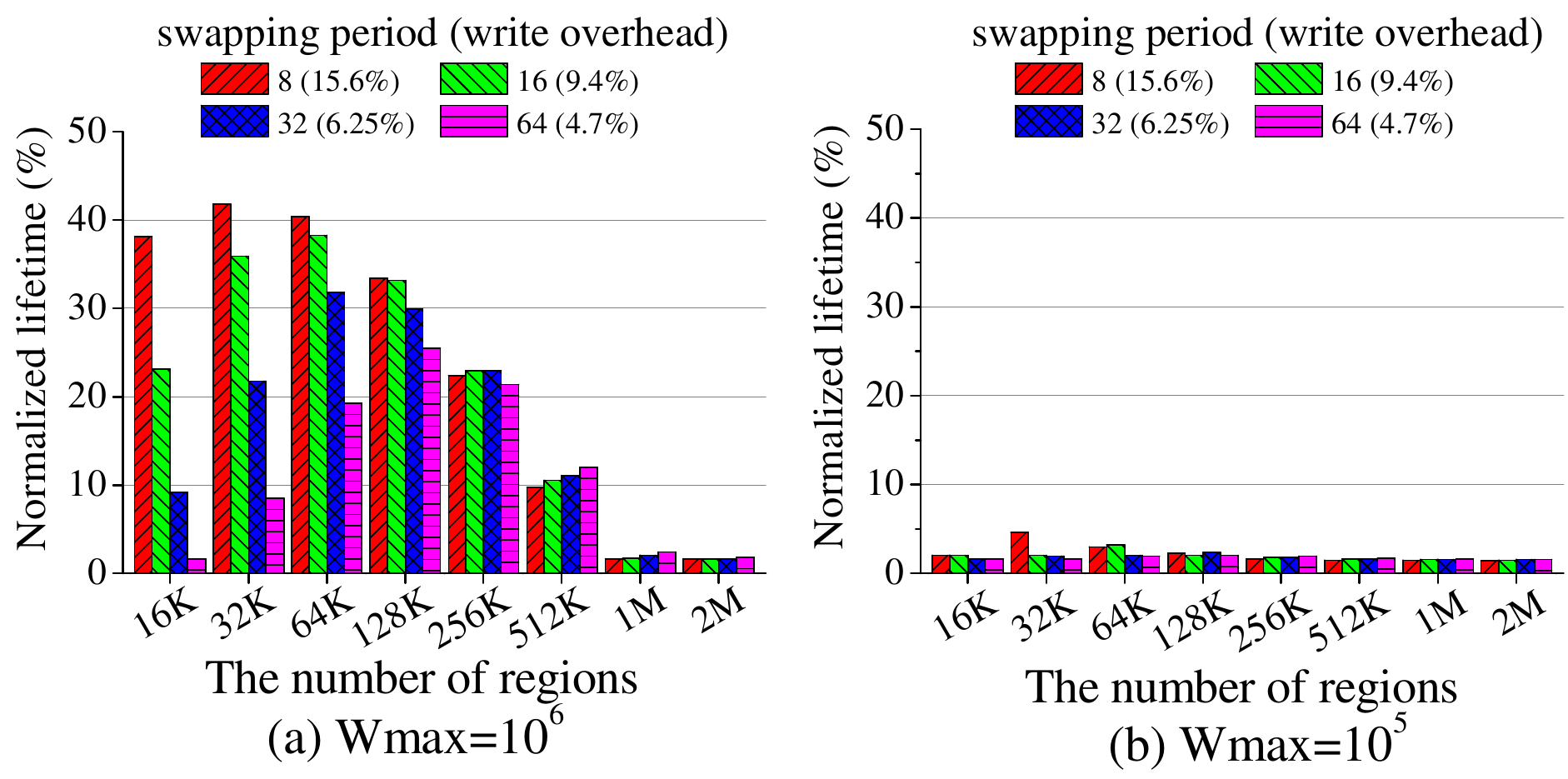}
\caption{The normalized lifetime of a 64GB NVM system with TLSR algorithm under the BPA program.}
\label{tlsr}
\vspace{-0.3cm}
\end{figure}

To quantitatively analyze and understand the security vulnerability problem of the existing wear-leveling algorithms, we conduct an experiment to evaluate the lifetime of MLC-based NVM devices under the Repeated Address Attack (RAA) ~\cite{qureshi2011practical} and Birthday Paradox Attack (BPA) ~\cite{seznec2010phase}. RAA is an attack program that writes data to the same address repeatedly. BPA aims to randomly select logical addresses and repeatedly write to each one precisely until being remapped to another physical address.

{\textbf{1) RAA risk for Segment Swapping and RBSG.}}  Since the Segment Swapping does not change the inter-segment offset address, the RAA programs are written back to the physical memory lines with the same offset address among the segments. These memory lines are worn out at the early stage. The RBSG, which adopts a static address mapping algorithm, fails to defend against the RAA program since the attacked physical address cannot be migrated to the entire address space. The attacked region then receives an extremely, disproportionally large number of writes, and fails in several hours. Therefore, we do not evaluate the experiments on RBSG and Segment Swapping as they are obviously unsuitable for large-capacity MLC-based NVM.

Since TLSR, PCM-S and MWSR algorithms can effectively migrate the attacked memory lines to the entire space to resist RAA program, we use the BPA program to evaluate the lifetime of MLC-based NVM system. We simulate a 64GB MLC-based NVM with 32 2GB banks and 256M memory lines, including 4M spare lines to tolerate some worn-out memory lines to prevent it from early failures. A line fails when its write count reaches its write limit. The NVM fails when there are not enough spare lines to replace the failing lines. With an assumed write limit of ${10}^{5}$ and ${10}^{6}$ for each cell~\cite{palangappa2017compex,xu2013understanding}, the ideal lifetime of this NVM system can be derived to be 2.5 months and 25 months respectively with 1GBps write traffic. For the TLSR, the outer-level swapping period is fixed at 32 and the inner-level swapping period varies from 8 to 64, while the accumulated number of regions increases from 16K to 64M.
.

{\textbf{2) Lifetime shortening for TLSR.}} Fig. \ref{tlsr} shows the normalized lifetime (i.e., to the ideal lifetime) of an MLC-based NVM system with the TLSR algorithm under the BPA program. The experimental results indicate that the lifetime of the MLC-based NVM system shows a trend from increase to decrease with the growing of the number of regions. When the number of regions is 32K, the MLC-based NVM system achieves the best lifetime, which means that the write counts of the memory lines within and among the regions are well balanced. In addition, the swapping period has a greater impact on NVM lifetime. When the number of regions is small (i.e., a region contains many memory lines), the low swapping period can increase the number of data exchanges, and thus achieves better wear leveling. However, when the number of regions is large, the low swapping period improves NVM lifetime slightly. On the contrary, the low swapping period greatly increases the number of data exchanges, incurring many extra writes and thus reducing the lifetime of NVM system. Furthermore, with the increase of the number of regions, the write distribution among the regions is more uneven, and the regions with heavily writes are easily worn out.

As shown in Fig. \ref{tlsr}(a), the best lifetime of the NVM system is 42\% of the ideal lifetime when the number of regions is 32K and the swapping period is equal to 8. However, this comes at the cost of a 15.6\% extra write overhead, which results in a severe performance degradation. As the swapping period increases to 32, the write overhead decreases to 6.25\%, but the system lifetime decreases to no more than 25.4\% of the ideal lifetime with the configuration of 64K regions. When the cell endurance decreases to ${10}^{5}$, the NVM system using TLSR lasts for 4.6\% of the ideal lifetime, as shown in Fig. \ref{tlsr}(b). Thus, TLSR is not competent for the work of wear leveling with MLC-based NVM.

\begin{figure}[t]
\vspace{-0.3cm}
  \centering
    \subfloat[$10^6$ endurance]{
    \label{fig:direct-parallel-b}
    \includegraphics [width=0.225\textwidth]{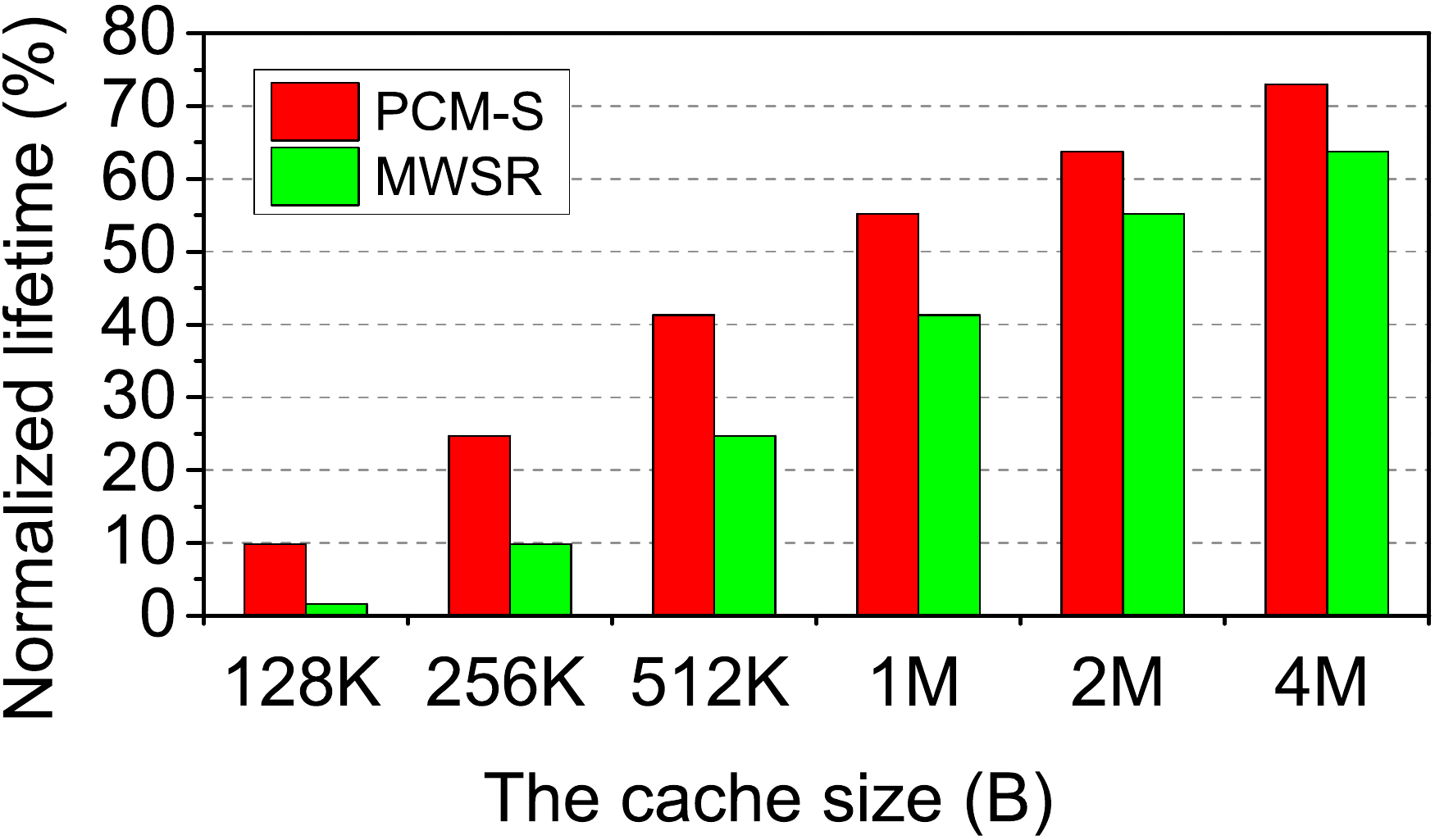}
    }
    \subfloat[$10^5$ endurance]{
    \label{fig:direct-parallel-c}
    \includegraphics [width=0.225\textwidth]{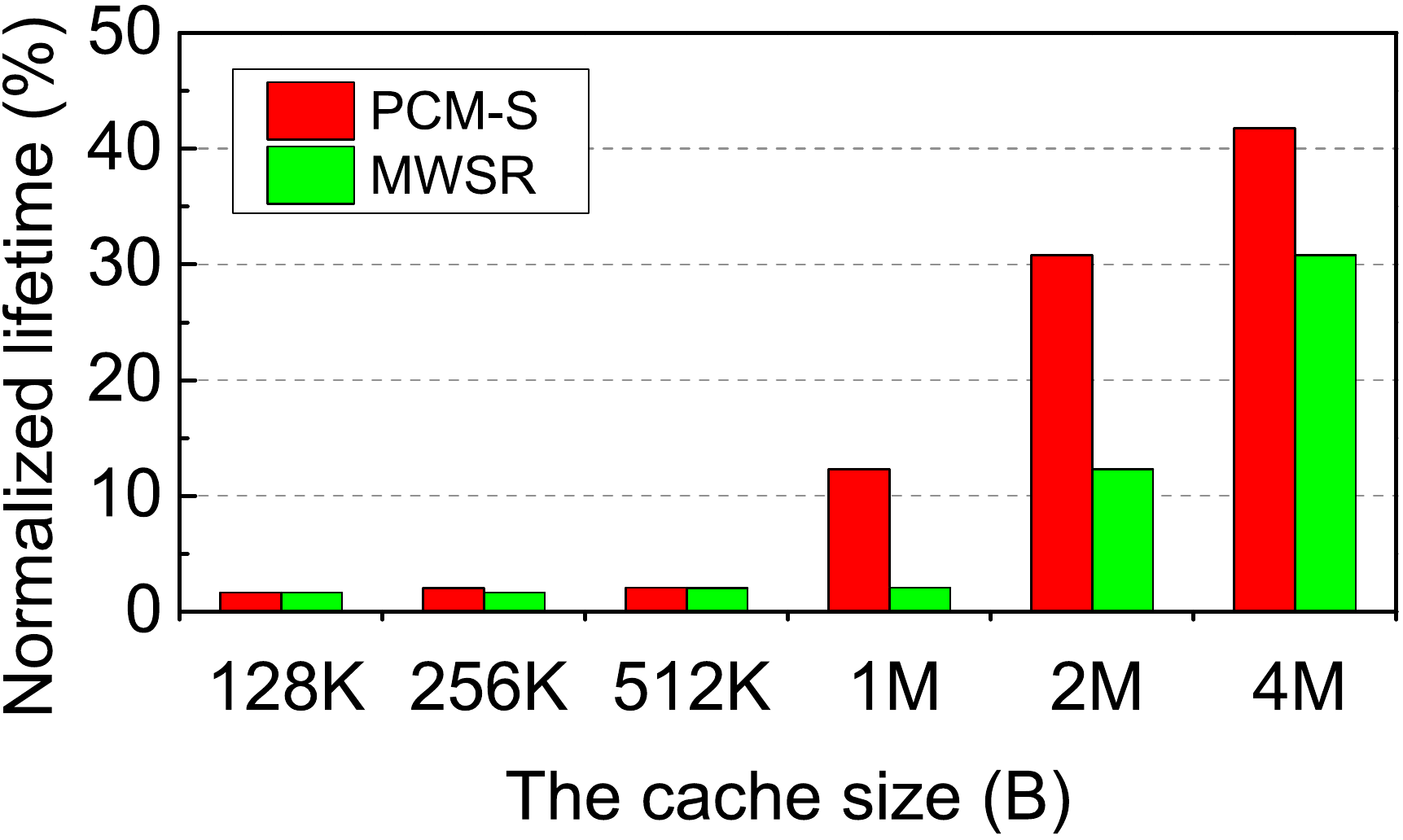}
    }
    \caption{\label{pcm_s} The normalized lifetime of a 64GB NVM system with PCM-S and MWSR under the BPA program.}

\end{figure}

{\textbf{3) Significant on-chip storage overhead for PCM-S and MWSR.}}
{Hybrid wear-leveling schemes need to store all address mappings in an on-chip cache. Specifically, PCM-S needs to record the physical address and internal offset of each logical region. MWSR needs to store two physical addresses (i.e., the physical addresses of the previous and current rounds), two offset addresses (i.e., the internal offsets of the previous and current rounds) and a write counter, for each logical region. Therefore, the space overheads of on-chip cache in PCM-S and MWSR algorithms are proportional to the number of regions. Using smaller wear-leveling granularity is able to increase the NVM lifetime but increases the number of regions and thus needs a larger on-chip cache.}
{We evaluate the NVM lifetime when PCM-S and MWSR are performed on MLC-based NVM with different on-chip cache sizes, as shown in Fig.~\ref{pcm_s}. We observe that PCM-S only achieves 72\% of ideal lifetime for the MLC-based NVM with $10^6$ endurance, and 41\% of ideal lifetime for the MLC-based NVM with $10^5$ endurance, even with a very large on-chip cache, i.e., 4MB. MWSR achieves the lower lifetime than PCM-S due to larger storage overhead of address mappings.}

{In summary, Segment Swapping and RBSG are vulnerable to RAA. TLSR causes significant NVM lifetime reduction. Hybrid wear-leveling algorithms including PCW-S and MWSR have the potential of achieving a high lifetime but cause significant on-chip storage overhead.}

\vspace{-0.3cm}
\section{Design and Implementation}

{To improve the NVM lifetime and reduce the on-chip storage overhead of hybrid wear-leveling algorithms, a naive solution called naive wear-leveling scheme (NWL) is to store all address mappings in the NVM and maintain recently-accessed mapping entries in an on-chip cache.
Nevertheless, the NWL often exhibits poor cache utilization under applications with substantial random access patterns, resulting severe system performance degradation due to long latency of accessing address mappings in NVM.}
To effectively address this problem, we propose a self-adaptive wear-leveling scheme, SAWL, to significantly improve the cache hit rate by dynamically and adaptively tuning the wear-leveling granularities at runtime based on the workload. The SAWL scheme enables the MLC-based NVM systems to attain high performance and long lifetime simultaneously. In what follows, we describe the tiered architecture and the self-adaptive wear-leveling scheme in detail.

{\subsection{An Architectural Overview}}

\begin{figure}[t]
\vspace{-1.5cm}
\centering
\includegraphics[width=0.5\textwidth]{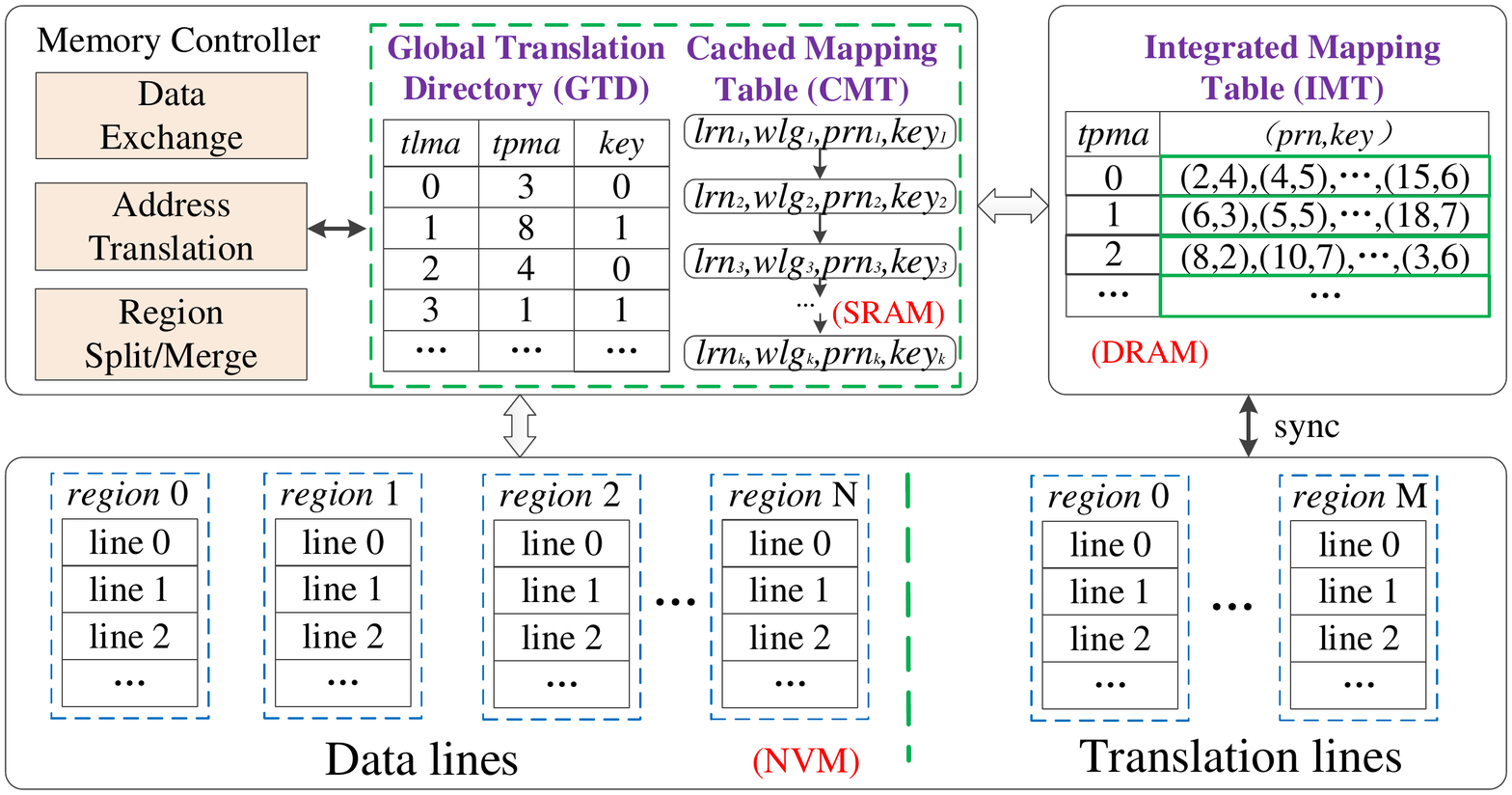}
\caption{The self-adaptive tiered wear-leveling architecture with NVM-based main memory.}\label{integrated}
\vspace{-0.5cm}
\end{figure}

SAWL is a tiered wear-leveling architecture consisting of a data exchange module, an address translation module and a region reconfiguration module, as shown in Fig. \ref{integrated}. The data exchange module is capable of implementing arbitrary hybrid wear-leveling algorithms. Since the address translation and region reconfiguration of PCM-S are relatively simple, we adopt PCM-S algorithm in data exchange module. The detailed data exchange algorithms are described in Section 2, and the relevant addresses, depending on their temporal and spatial properties, are stored in an Integrated Mapping Table (IMT), a Cached Mapping Table (CMT) and a Global Translation Directory (GTD), which are managed by the address translation module. 

SAWL uses translation lines to record the locations, in which the user data are actually stored. 
To prevent the translation lines from being worn out, the NVM system must independently perform hybrid wear leveling for the translation lines. Hence, a GTD table is needed to record the relationship between the logical translation line memory address (\emph{tlma}) and its physical counterpart (\emph{tpma}). The GTD table can be entirely stored in the SRAM due to its extremely low space overhead. In the meantime, to prevent the loss or corruption of the metadata (e.g., data stored in the CMT, GTD and IMT tables) due to power failures, the updated metadata are written back to the NVM devices. Within the long swapping period, the update operation is infrequent, which has negligible influence on NVM performance. How to ensure the crash consistency is important an challenging problem and has been discussed in ~\cite{liu2018crash,zuo2018secpm,wei2017transactional}, which is beyond the scope of this paper and we assume that there is a battery backup in memory controller to refresh metadata during power failure like existing schemes~\cite{liu2018crash,shin2017proteus}.

IMT records the relationship between a logical region number ($lrn$) and its corresponding physical region number ($prn$), where $lrn$ represents the $N$ Most Significant Bits (MSB) of the logical memory address and an $lrn$ can be mapped to any physical region. In addition, IMT records the offset parameter ($key$) of each region, through which we obtain the corresponding intra-regional physical address offset. The $lrn$ is implicitly indicated by IMT. Assuming a translation line in IMT contains 6 translation entries (determined by the size of translation entry), the first line contains $lrn0$ to $lrn5$ at the beginning. And after several translation line remapping, the first line may contain $lrn6k$ to $lrn6k+5$, where $k$ is an integer. We obtain the \emph{tpma} from GTD table using \emph{tlma} and finally get user data line information from IMT table. The size of the IMT table, e.g., tens to hundreds of megabytes, is proportional to the NVM capacity and too large to be entirely held in the memory controller. Therefore, the IMT table is stored in a reserved space of the NVM devices with its entries packed into memory lines that are called \emph{translation lines}, in contrast to the data lines that hold users' data. The entries are placed in an ascending order of the $lrn$ to facilitate easy address lookup. To alleviate performance degradation induced by long address translation latency, a naive scheme is to leverage DRAM or NVM to hold complete IMT table and a CMT table in the SRAM to buffer the recently-used IMT entries. The entries in CMT are organized in an LRU stack and a new entry cached from NVM will evict the least-recently-used entry in the CMT.

Moreover, we use a parameter, called \emph{wear-leveling granularity} ($wlg$), to represent the range of the address space covered by each entry. When SAWL changes the wear-leveling granularities, this parameter needs to be updated.

\subsection{Self-Adaptive Wear Leveling}

With the limited cache space, only a relatively small number of mapping entries can be held in the cache. When applications exhibit very weak locality and the requested addresses are sparsely dispersed over the entire address space, {the NVM system exhibits very poor cache hit rate and performance. To address this problem, we propose the SAWL scheme.}

\begin{figure}[t]
\centering
\includegraphics[width=0.42\textwidth]{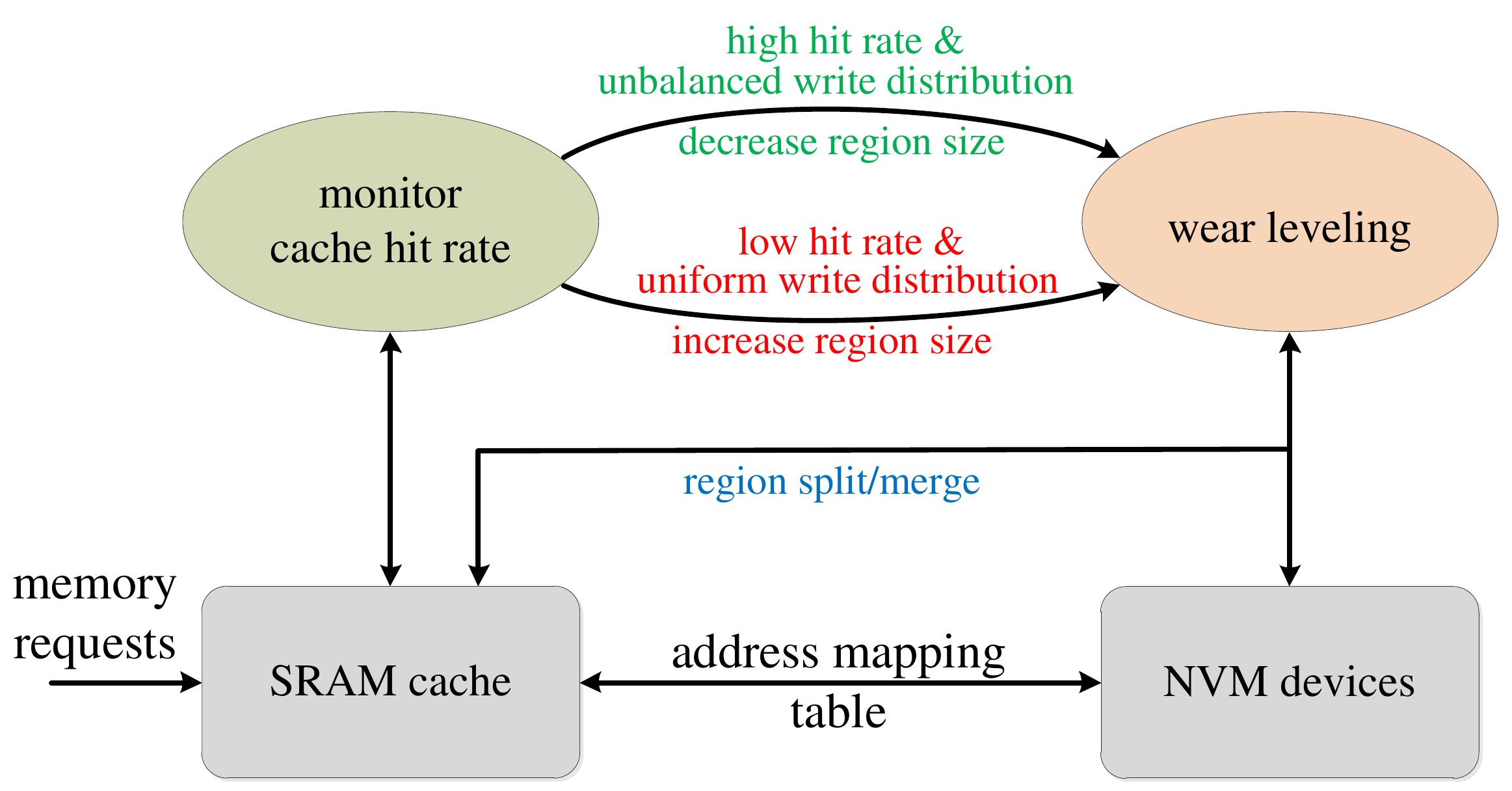}
\caption{An overview of self-adaptive wear leveling.}\label{SAWL}
\vspace{-0.5cm}
\end{figure}

Based on the experimental results shown in Fig.~\ref{pcm_s}, we observe that under a hybrid wear-leveling algorithm, the lifetime of an NVM system is generally positively correlated to the number of regions. In other words, the larger the number of regions is, the closer the NVM system approaches its ideal lifetime. However, with an increasing number of regions, the number of memory lines within a region is reduced. Hence, the address space covered by each of the Cached Mapping Table (CMT) entries, i.e., wear-leveling granularity, decreases accordingly. Since the number of CMT entries is fixed, the whole address space covered by the SRAM cache decreases, which reduces the cache hit rate.

To address this performance problem, the design goal of SAWL is to automatically tune the region size to improve NVM performance whenever the SRAM cache demonstrates poor hit rate under some applications, as shown in Fig. \ref{SAWL}. To achieve this goal, SAWL carries out a region-merge operation to merge two or more regions into a single larger region, thus allowing an Integrated Mapping Table (IMT) entry to cover more addresses and increasing the wear-leveling granularity. On the other hand, since a coarse wear-leveling granularity reduces wear leveling, SAWL counters this by carrying out a region-split operation to divide a large region into multiple smaller regions when the cache hit rate continues to climb beyond a predefined threshold and the hits have become severely unbalanced within the region. In addition, the NVM lifetime can be used as an indicator to tune wear-leveling granularities. However, the lifetime is difficult to measure during runtime. In general, the lifetime is calculated by running many requests until the NVM cell is worn out. Since the cache hit rate is easy to capture, we adopt the indicator of the cache hit rate which shows the performance decrease of NVM system.

{\textbf{1) Region-merge operation.}} To perform the region-merge operation, SAWL first picks out the physical location for the new region, in which the physical location is mapped by one of non-merged logical locations to avoid choosing the physical locations that have been occupied by other already merged regions. Then, SAWL chooses the closest non-merged logical location and merges them.

\begin{figure}[t]
\centering
\includegraphics[width=0.47\textwidth]{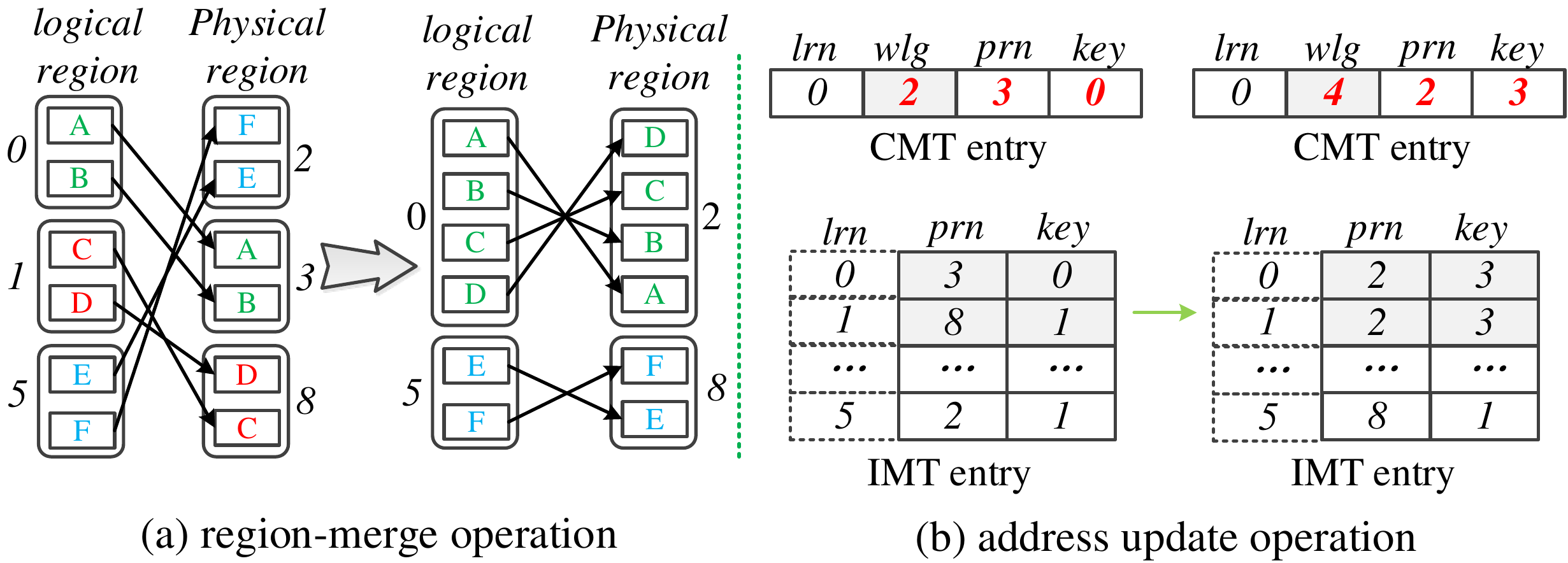}
\caption{An example of the region-merge and address update operations. lrn is implicitly indicated by IMT. }\label{regionmerging}
\vspace{-0.3cm}
\end{figure}

SAWL further swaps the data of the new region with the data of the new location, ensuring that the logical addresses and their physical counterparts of the memory lines within the newly merged region satisfy the algebraic mapping. Finally, we update the address-mapping table on the NVM and the relevant CMT entries on the SRAM. Fig. \ref{regionmerging} depicts an example of the region-merge operation. As shown in Fig. \ref{regionmerging} (a), there are three logical regions, e.g., ${lrn}_{0}$, ${lrn}_{1}$ and ${lrn}_{5}$, which are mapped to ${prn}_{3}$, ${prn}_{8}$ and ${prn}_{2}$, respectively. To merge ${lrn}_{0}$ and its closet logical neighbour ${lrn}_{1}$ into one super region, we pick out a large physical space for the newly merged region (e.g., ${prn}_{2}$ and ${prn}_{3}$). We then move out the lines E and F from ${prn}_{2}$ (the data can be temporarily stored on memory controller or DRAM cache). We also migrate the lines C, D of ${lrn}_{1}$ to ${prn}_{2}$, and rotate all the memory lines within  ${prn}_{2}$ and ${prn}_{3}$ to ensure addresses of the logical memory lines within the two regions satisfy the algebraic mapping function. Finally, we write back the data of lines E and F to ${prn}_{8}$ and update the corresponding entries in IMT and CMT tables, as shown in Fig. \ref{regionmerging} (b). After this, the $wlg$ parameter of ${lrn}_{0}$ is changed to 4, which means the ${lrn}_{0}$ entry covers four memory addresses at present. Since ${lrn}_{0}$ and ${lrn}_{1}$ belong to the same large region, the physical region address and internal offset in IMT are identical. On the address translation, the NVM obtains the real wear-leveling granularity of a region based on the number of adjacent regions which have the same address information.

\begin{figure}[t]
\centering
\includegraphics[width=0.49\textwidth]{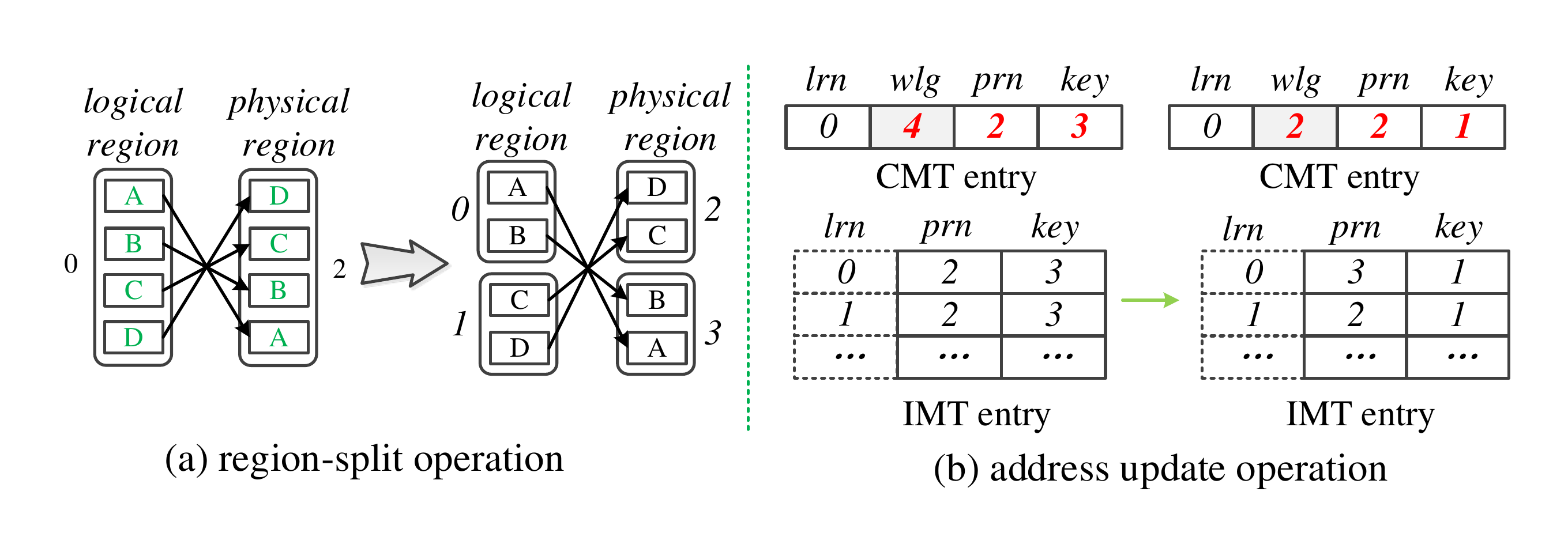}
\vspace{-0.3cm}
\vspace{-0.3cm}
\caption{An example of the region-split and address update operations. lrn is implicitly indicated by IMT. }\label{regionspliting}
\vspace{-0.3cm}
\end{figure}

{\textbf{2) Region-split operation.}}  In contrast to the region-merge operation, the region-split operation splits a large region into multiple smaller regions by migrating the memory lines within the large region. More specifically, if we use the XOR operation to conduct address mapping, there is no need to migrate the memory lines within the old large region since the XOR operation makes the memory lines within each post-split smaller region contiguous in the physical space. We only need to update the address-mapping table and the CMT entries, and the memory lines have already satisfied the algebraic mapping function. Fig. \ref{regionspliting} describes an example of a simple region-split operation. As shown in Fig. \ref{regionspliting} (a), the large region ${lrn}_{0}$ is split into two sub-regions (${lrn}_{0}$ and ${lrn}_{1}$). Given that the memory lines in ${lrn}_{0}$ and ${lrn}_{1}$ are mapped to the same physical sub-regions, there is no need to migrate the memory lines if we keep the original mapping relationship. Since the wear-leveling granularities of ${lrn}_{0}$ and ${lrn}_{1}$ changes, we only update the relevant entries in IMT and CMT tables. The new physical address of the sub-regions is obtained by the region address XORing with the most significant bit (MSB) of the offset parameter, e.g., the keys of ${lrn}_{0}$ and ${lrn}_{1}$. For example, the physical address of ${lrn}_{0}$ is calculated by the $2 \oplus 1$, where '1' denotes the MSB of the old key of ${lrn}_{0}$. The new keys is achieved by the least significant bits (LSB) of the old key, e.g., the old key of ${lrn}_{0}$ and ${lrn}_{1}$ are 3 ('11') and its LSB is '1' as shown in Fig. \ref{regionspliting} (b). After the region-split completes, the ${lrn}_{0}$ and ${lrn}_{1}$ do not belong to a large region because their physical addresses are different. In contrast to region-merge operations, the overhead of region-split operations is extremely low.

\begin{figure}[t]
\centering
\includegraphics[width=0.38\textwidth]{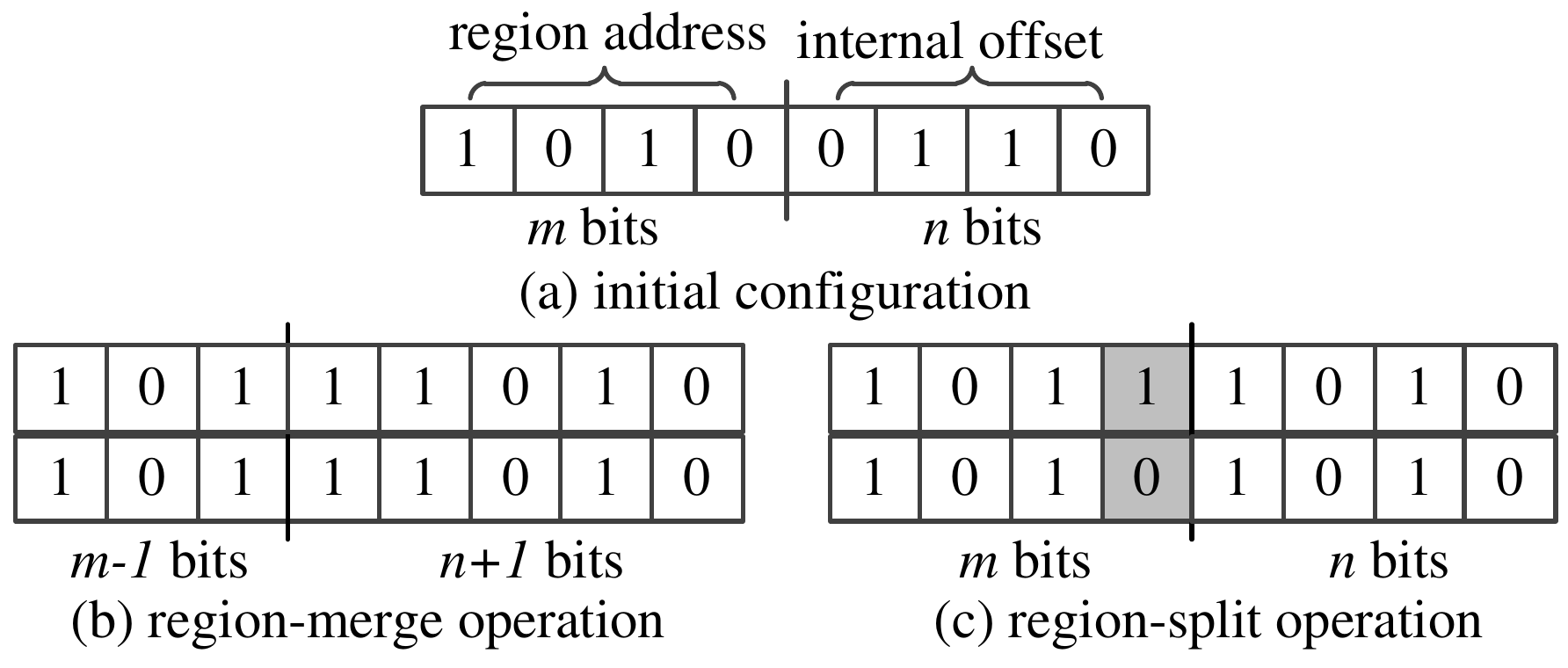}
\caption{An example of size scaling for IMT entries.}\label{imtupdating}
\vspace{-0.3cm}
\end{figure}

To make the region-split operation efficient, we employ two registers to record the cache hit counts of the first and the second half of the CMT entries queue, respectively. Since the entries in CMT are organized in an LRU stack, usually the hit count of the first sub-queue is larger than that of the second one. If the first one is far larger than the second one, it means that the addresses in the second sub-queue are rarely accessed, and splitting the region is beneficial. Otherwise, the current region size is of a satisfactory wear-leveling granularity.

To avoid performing the region-split and region-merge operations too frequently, SAWL tunes the region size only when the cache hit rate stays over the high threshold or below the low threshold for certain number of requests. Considering the relatively large region-merge overhead, we only merge the cached regions rather than all the regions in the entire memory.

{\textbf{3) Implementation details.}} For the concrete implementation, the NVM systems use the reserve space to store the IMT table. The capacity of IMT is determined by an initial wear-leveling granularity (P), i.e., the number of IMT entries equals ${M/P}$, where $M$ denotes the number of lines in the entire memory. In the working process, the size of IMT doesn't change. Otherwise, the address migration incurs massive space overhead, and the address translation becomes extremely complex. Fig. 
\ref{imtupdating} shows an example of address update for IMT. Each IMT entry records the address information (including the region address and offset parameter) according to the initial configuration. For example, $m$ bits keep the region address, and $n$ bits record the offset parameter. The sum of $m$ and $n$ is fixed and determined by the NVM capacity, i.e., $m+n=log_{2}^{M}$. After region-merge operation completes, the region size increases and the number of regions decreases. Thus we use a small amount of bits to record the region address and leverage more bits to record the offset parameter. As shown in Fig. \ref{imtupdating} (b), NVM uses $m$$-$$1$ and $n$$+$$1$ bits to record the region address and offset parameter, respectively. To indicate the sub-regions belonging to a large region, their address information is identical. After region splitting completes, the number of regions increases and more bits are required to record the region address, while less bits are used to keep the intra-regional offset, as shown in Fig. \ref{imtupdating} (c). In addition, the address information of the adjacent regions is different. It is worth noting that the minimum wear-leveling granularity cannot be smaller than the initial configuration, because the shortened wear-leveling granularity will significantly increase the size of IMT table and the NVM does not have sufficient reserved space to store the increased address-mapping table. The region-split and region-merge operations result in a dynamic tuning of the wear-leveling granularities. 
Given that the number of adjacent regions that have same address information is $n$, the real wear-leveling granularity (Q) is calculated by the formula of $Q = n \times P$.

\subsection{{Adaptive} Address Mapping Algorithm}

\begin{figure}[t]
\centering
\includegraphics[width=0.41\textwidth]{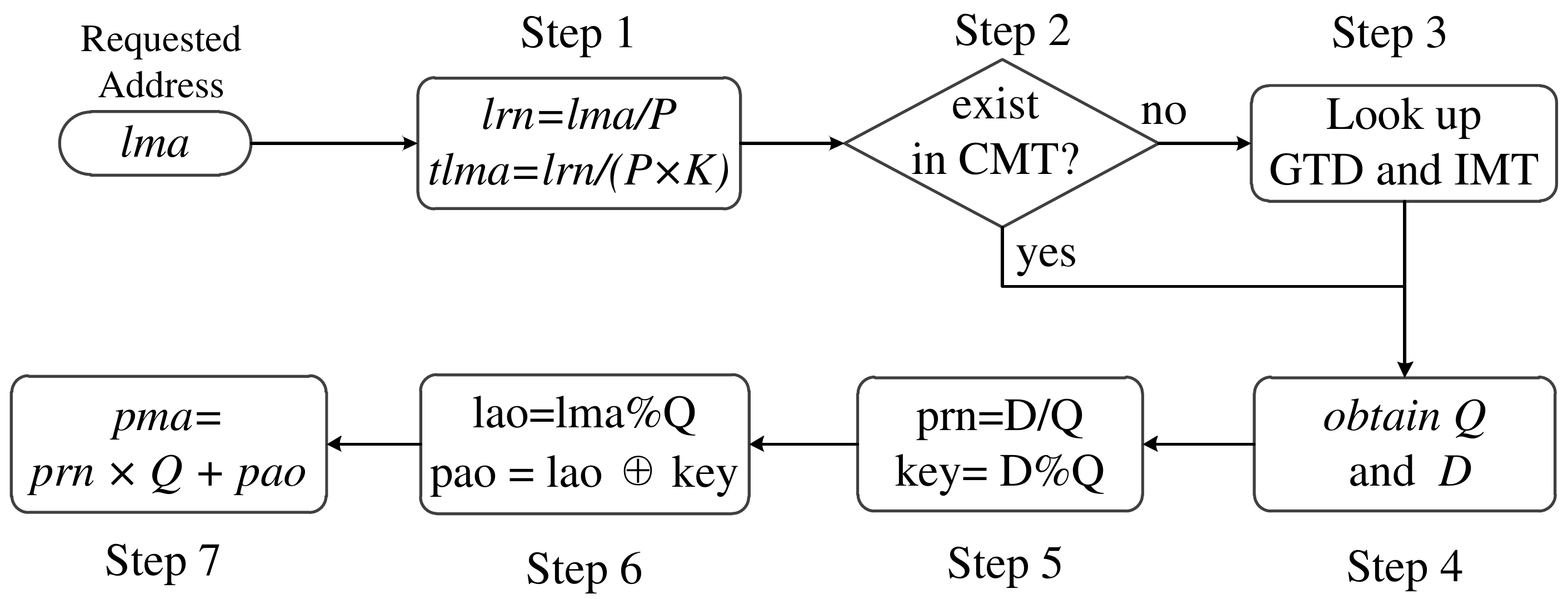}
\caption{The workflow of the address translation.}
\label{address_translation}
\vspace{-0.5cm}
\end{figure}

The wear-leveling process makes the relationship between a logical address and its physical counterpart dynamic. When a request arrives at the memory controller, the requested address must be translated into a physical address that is used to access the underlying NVM devices. This address translation in SAWL is facilitated by the Global Translation Directory (GTD), Cached Mapping Table (CMT) and Integrated Mapping Table (IMT). The 6-step workflow of the address-mapping algorithm is shown in Fig. \ref{address_translation}.
SAWL first computes the logical region number ($lrn$, $lrn= lma/{P}$) according to the given logical memory address ($lma$), and then obtains the logical translation line address ($tlma$, $\frac{lrn}{P \times K}$), where P represents the initial wear-leveling granularity and K is the number of entries within a translation line, which is 6 in our design (Step 1). The variable $lrn$ is used to check the CMT table to see if the translation entry is cached. If yes, the values of the real wear-leveling granularity (Q), address information ($D$, the combination of the physical region number and offset parameter) are obtained by accessing the SRAM cache  (Step 2). If it is a miss, the physical translation line addresses ($tpma$) value of the corresponding translation line is obtained from the GTD table. Using the \emph{tpma} value, the translation line can be read from IMT table in DRAM or NVM devices and placed at the top of the LRU stack (Step 3). From this line, the $Q$ and $D$ values of this requested address are found (Step 4). For example, if the $lrn$ we compute is $6k+m$, the $m$th entry in this line is needed. The physical region number ($prn$) and the offset parameter ($key$) are obtained based on the formula of $prn=\frac{D}{Q}$ and $key= D\%Q$, respectively (Step 5). Thus, the logical address offset ($lao$) and physical address offset ($pao$) are $lao=lma\%Q$ and $pao = lao \bigoplus key$ (Step 6). Finally, SAWL obtains the physical memory address ($pma$) by combining the $prn$ and $pao$ using $pma=prn \times Q + pao$ (Step 7).

Address translation leads to extra access latency to main memory. The main overhead includes looking up the CMT, GTD and IMT tables, respectively. In general, the access latency to the CMT and GTD tables would be about 5 ns due to their residence in SRAM, while a DRAM/NVM read operation is at least 50 ns. For the CMT table, the entries cached on SRAM are organized in the LRU list.  
Given that the SRAM query consumes 3ns, we hence set 5ns on average for address translation latency. Our SAWL dynamically tunes the wear-leveling granularities (i.e., region size) to increase the number of cached addresses, which improves the cache hit rate and I/O performance significantly.

\vspace{-0.3cm}

\section{Performance Evaluation}


\subsection{Methodology}

In our experiments, we use the Gem5 simulator \cite{binkert2011gem5} to evaluate various wear-leveling schemes and NVMain~\cite{poremba2015nvmain} to examine the lifetime and cache hit rate in a time-efficient way. We evaluate state-of-the-art hybrid wear-leveling algorithms, including the basic non-tiered architecture (BWL), {i.e., PCM-S and MWSR,} naive tiered architecture (NWL) and compare them with our SAWL algorithm on the tiered architecture.

In the following experiments, the initial wear-leveling granularity of BWL, NWL and SAWL is set to 4 memory lines to ensure that the lifetime MLC-based NVM systems lasts for a long time under the worst-case attacks. In addition, we use NWL-4 and NWL-64 to respectively represent the naive wear-leveling algorithm on the tiered architecture with a region consisting of 4 and 64 memory lines respectively. For the SAWL scheme, the lowest region-merge threshold is set to 90\% based on the experimental observation that the cache hit rate of 90\% marks a turning point below which the performance of NVM system decreases significantly. For the region-split operation, the highest cache-hit-rate threshold is set to 95\%, because the performance evaluation indicates that the wear-leveling algorithm has slightly impact on NVM performance within the boundary. The SAWL algorithm automatically tunes the wear-leveling granularities when the cache hit rate is above or below this threshold for a long time. Moreover, if the hit ratio of the first queue OR the hit ratio of the second queue $\geq$ 99\%, the NVM system splits the region for endurance, thus avoiding the decrease of cache hit rate after region-split completes.

To evaluate the performance of NVM system under general applications, we use 14 representative applications from the SPEC CPU2006 suite \cite{gove2007cpu2006}, which contain high memory accessing frequency with at least 100 million read/write requests in each application. These applications have been widely used in existing lifetime analysis experiments \cite{qureshi2009enhancing,zhou2009durable,jiang2011lls}. We perform evaluations by executing the benchmark in rate mode, where all the eight cores execute the same benchmark \cite{young2015deuce}.

\subsection{Parameter Training via Sensitivity Study}

The key to SAWL is to dynamically adjust the region size, or wear-leveling granularities, by applying a combination of region-merge and region-split operations based on the workload behaviors that are monitored using the observed runtime cache hit rate. To accurately capture the runtime cache hit rate and adjust the region size in a reliable and cost-efficient way, SAWL relies on two critical parameters, the \emph{size of the observation window} for capturing runtime cache hit rate and the \emph{size of the settling window} for reliable and efficient region-size adjustment. In what follows we first define these parameters and then experimentally determine their values.

\begin{figure}[ht]
\centering
\includegraphics[width=0.45\textwidth]{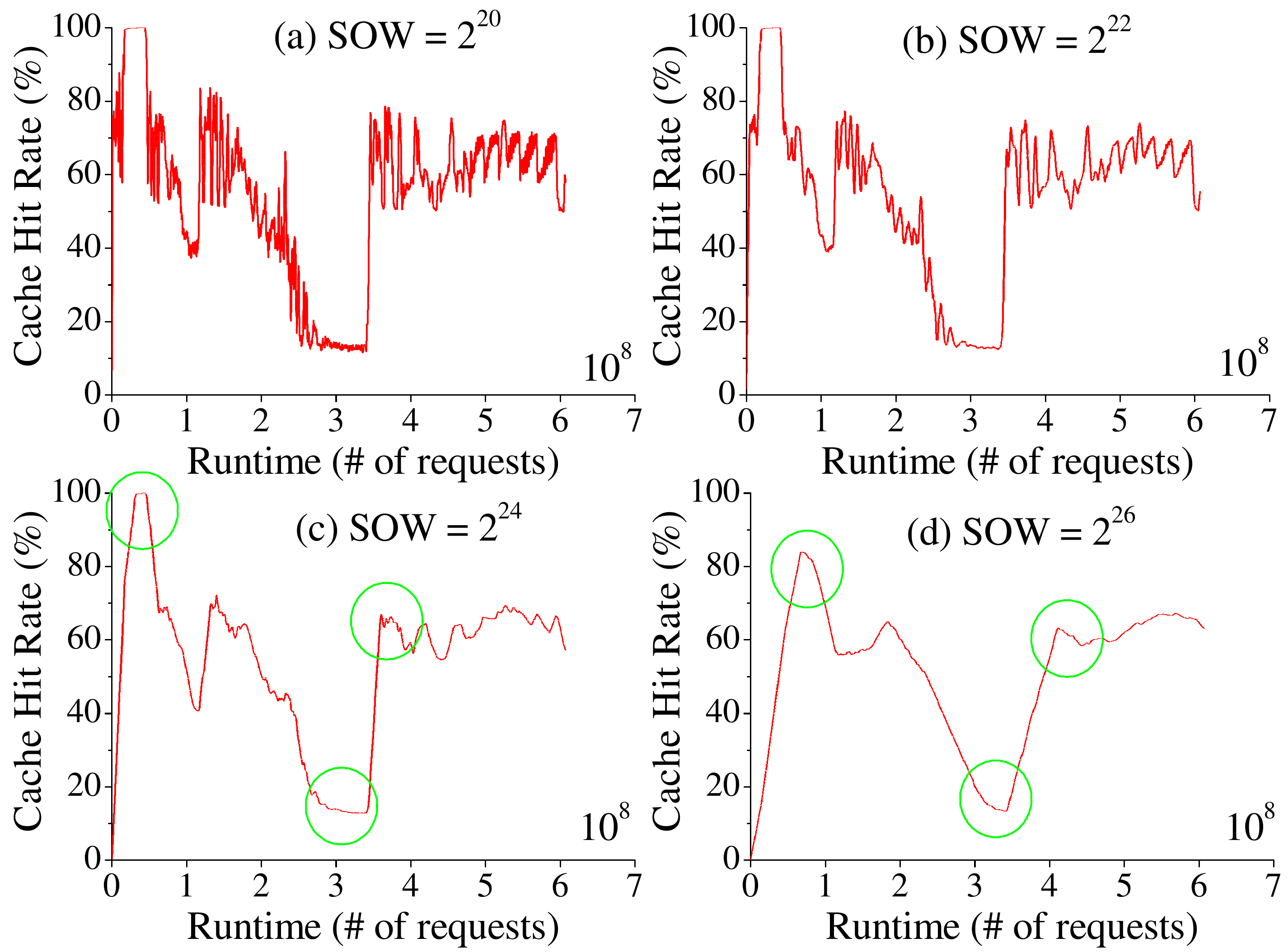}
\caption{Cache hit rate as a function of runtime obtained from different observation window sizes $SOW$ when running the SPEC CPU2006 $soplex$ benchmark in a 512KB cache.}
\label{observewindow}

\end{figure}

\begin{figure}[t]
\centering
\includegraphics[width=0.45\textwidth]{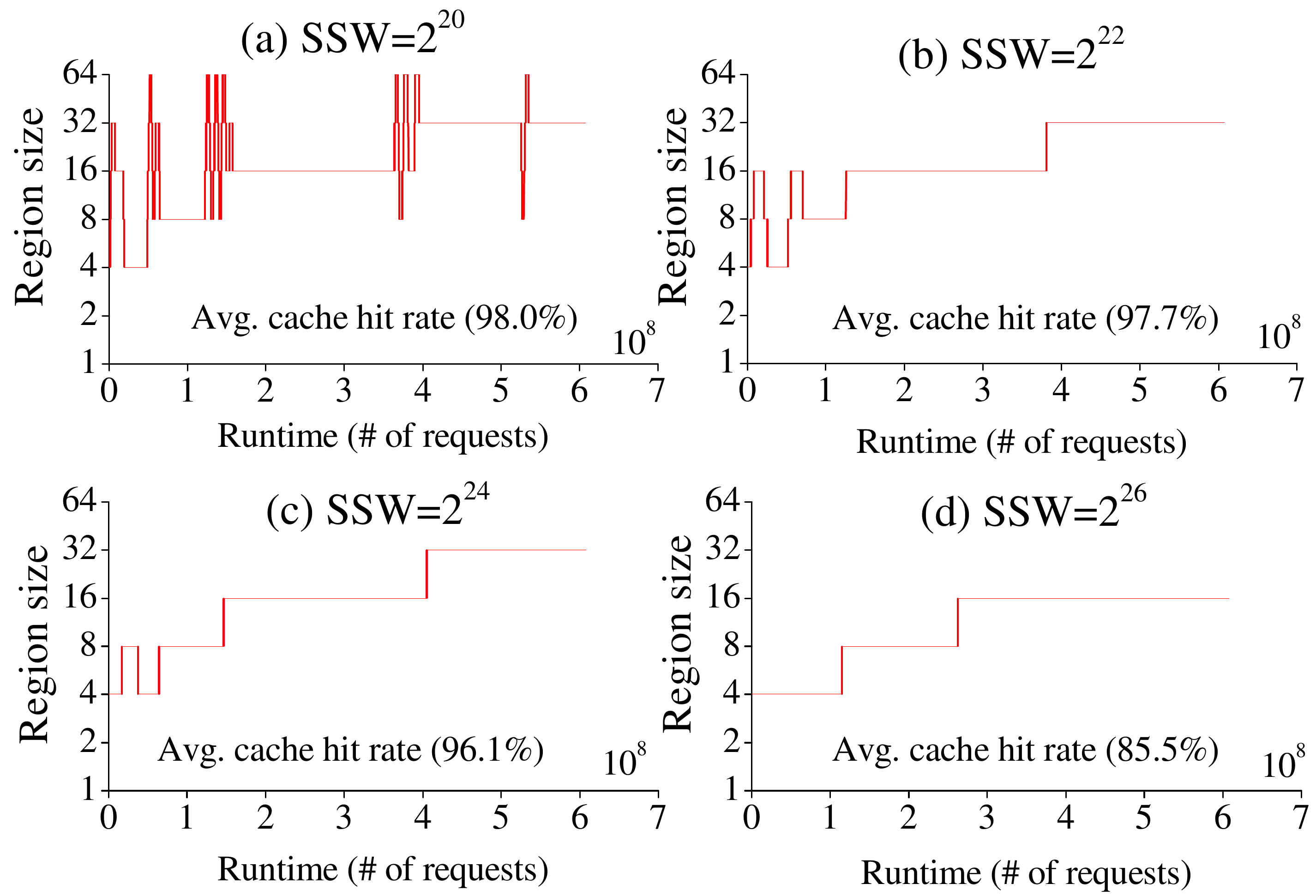}
\caption{The region size adjustments as a function of the runtime with different settling window sizes under the $soplex$ benchmark.}
\label{Continuous Window}
\vspace{-0.3cm}
\vspace{-0.2cm}
\end{figure}

\begin{figure*}[t]
\centering
\includegraphics[width=0.29\textwidth]{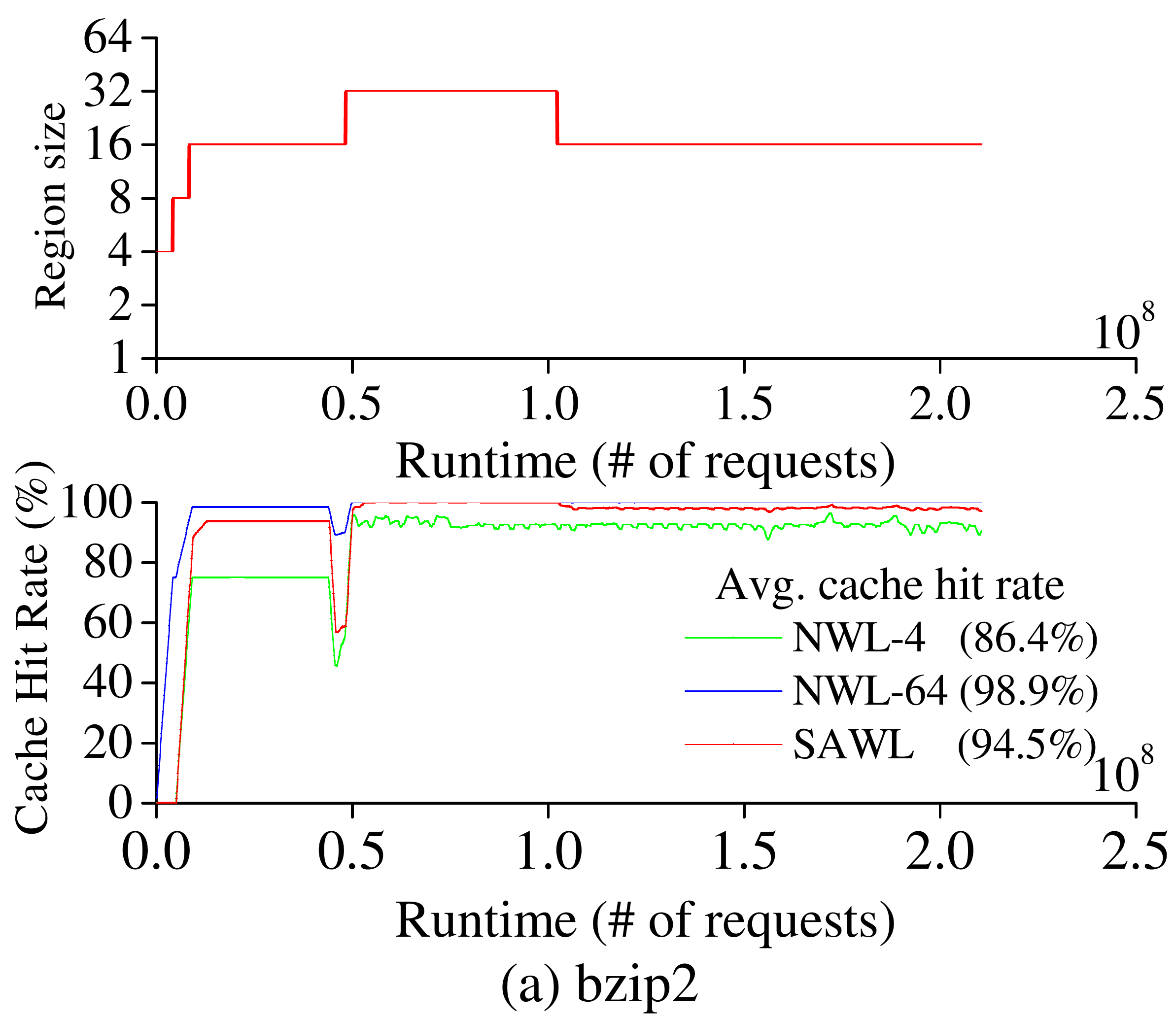}
\hspace{6px}
\includegraphics[width=0.29\textwidth]{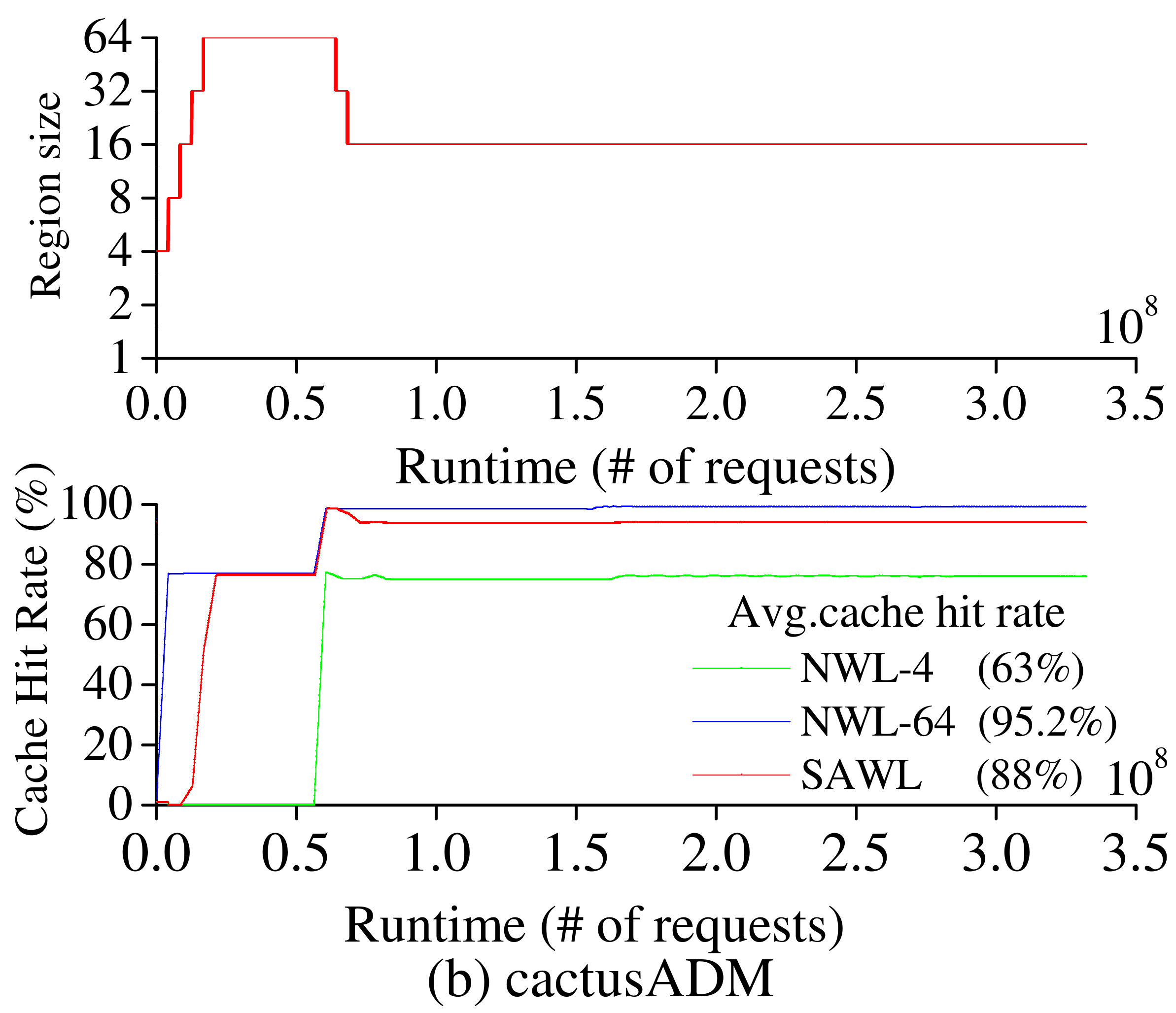}
\hspace{6px}
\includegraphics[width=0.29\textwidth]{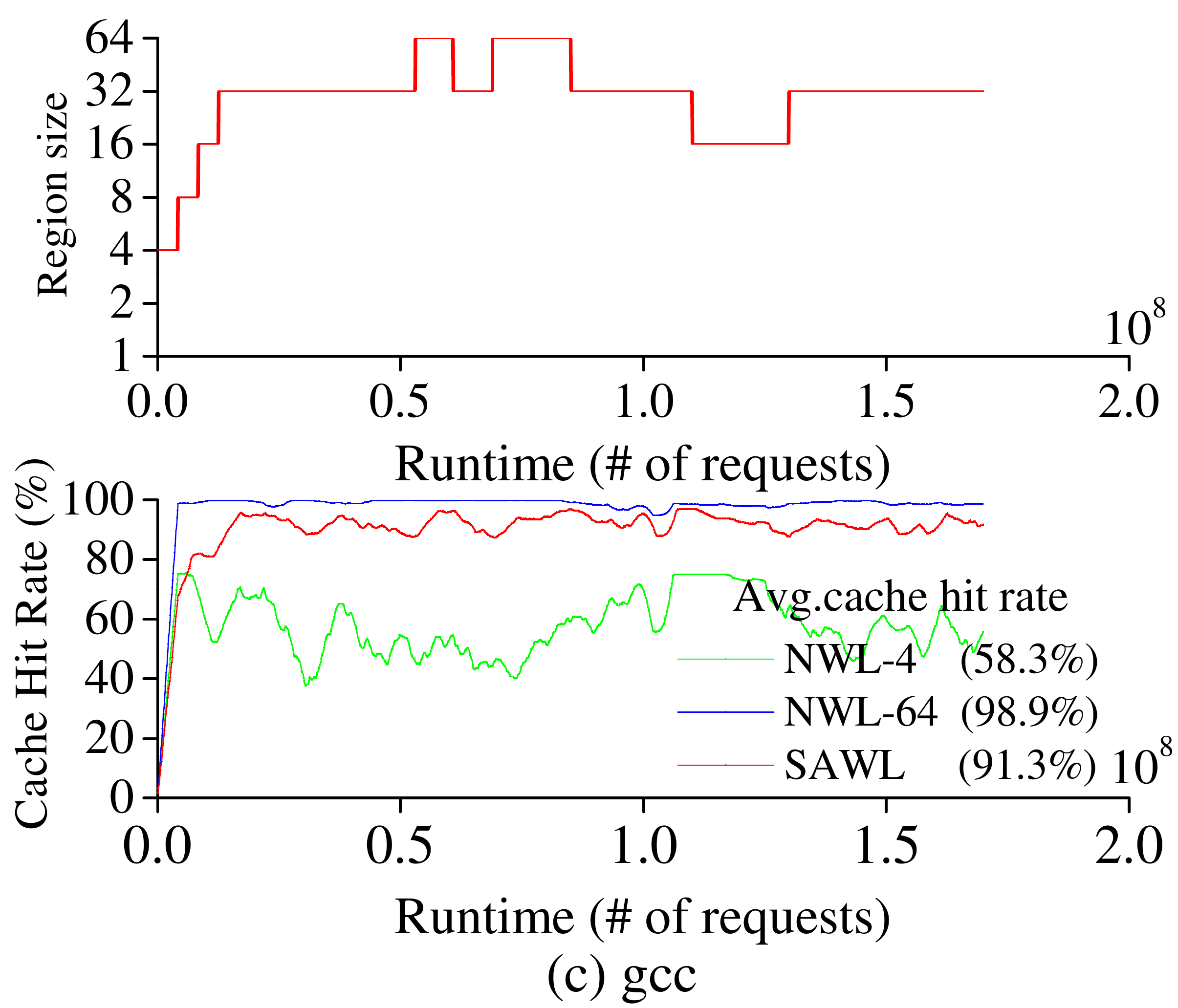}
\caption{The runtime hit rates and region size adjustments under the three representative benchmarks.}
\label{various-SAWL}
\vspace{-0.3cm}
\end{figure*}

{\textbf{1) Observation Window Size.}} SAWL measures the current runtime cache hit rate by calculating the percentage of memory access requests that hit the cache out of a certain total number of requests observed, including the most recent one. This total number $SOW$ of observed requests is called the \emph{size of the observation window}. We measure the runtime cache hit rate every $100,000$ requests as it is not very sensitive for the accuracy of measurement according to our experiments. However, our experiments revealed that $SOW$ is a sensitive parameter for the accuracy of the sampled cache hit rate. To find an optimal value for $SOW$, we examine how the sampled cache hit rate changes with the size of the observation window.

Fig. \ref{observewindow} shows the cache hit rates of different sizes of the observation window as a function of \emph{runtime}, which is defined by the total number of requests issued thus far, when running the SPEC CPU2006 $soplex$ benchmark in a 512KB cache. Specifically, as shown in Fig. \ref{observewindow}(a), when the window size is ${2}^{20}$, the cache hit rate fluctuates significantly causing SAWL to adjust the region size too frequently to be efficient. And as the observation window size ($SOW$) increases, the sampled cache hit rate becomes less fluctuating and more stable, which brings SAWL to miss the important time points, in these points, the SAWL needs to split or merge regions, as indicated by the small green circles in Fig. \ref{observewindow} (c) and (d). Consequently, we choose $2^{22}$ as the size of observation window.

{\textbf{2) Settling Window Size.}} SAWL waits for a certain number of requests to ensure that the cache hit rate of the observed runtime is sufficiently stable so as to avoid unnecessary or frequent region adjustments. This waiting period is called the \emph{settling window} and the number of requests to wait is called the \emph{size of the settling window} ($SSW$). Fig. \ref{Continuous Window} shows the adjustments of region size as a function of the runtime (i.e., the number of requests) with different $SSW$ values under the $soplex$ workload. Specifically, as shown in Fig. \ref{Continuous Window}(a), a small settling window size, i.e., $2^{20}$, results in frequent region size adjustments and incurs high write overhead. On the contrary, Fig. \ref{Continuous Window}(d) indicates that a large settling window size leads to SAWL to fail to sufficiently adjust the region size and obtain high performance since SAWL misses important time points of splitting and merging regions. In fact, the cache hit rate decreases to $85.5\%$. As a result, we argue that the settling window sizes in Fig. \ref{Continuous Window} (b) and (c) are much better. By training the parameters in Fig. \ref{observewindow} and \ref{Continuous Window}, we experimentally determine the best $SOW$ and $SSW$ values are both $2^{22}$.

In order to validate the efficiency and effectiveness of the values of $SOW$ and $SSW$ determined experimentally above, we evaluate the average cache hit rates of SAWL under the three representative benchmarks of $bzip2$, $cactusADM$ and $gcc$ respectively. As shown in Fig. \ref{various-SAWL}, the average cache hit rates of the three workloads are 94.5\%, 88\% and 91.3\%, respectively, which are close to those of NWL-64. SAWL improves the hit rates via increasing the region size when the hit rate becomes too low. Furthermore, the average region size of SAWL is about 16 memory lines in all workloads, which means that the BPA lifetime of NVM is about 20 months even under the worst-case workload.

\subsection{NVM Lifetime}

\begin{figure}[t]
  \centering
    \subfloat[$10^6$ endurance]{
    \label{fig:direct-parallel-b}
    \includegraphics [width=0.225\textwidth]{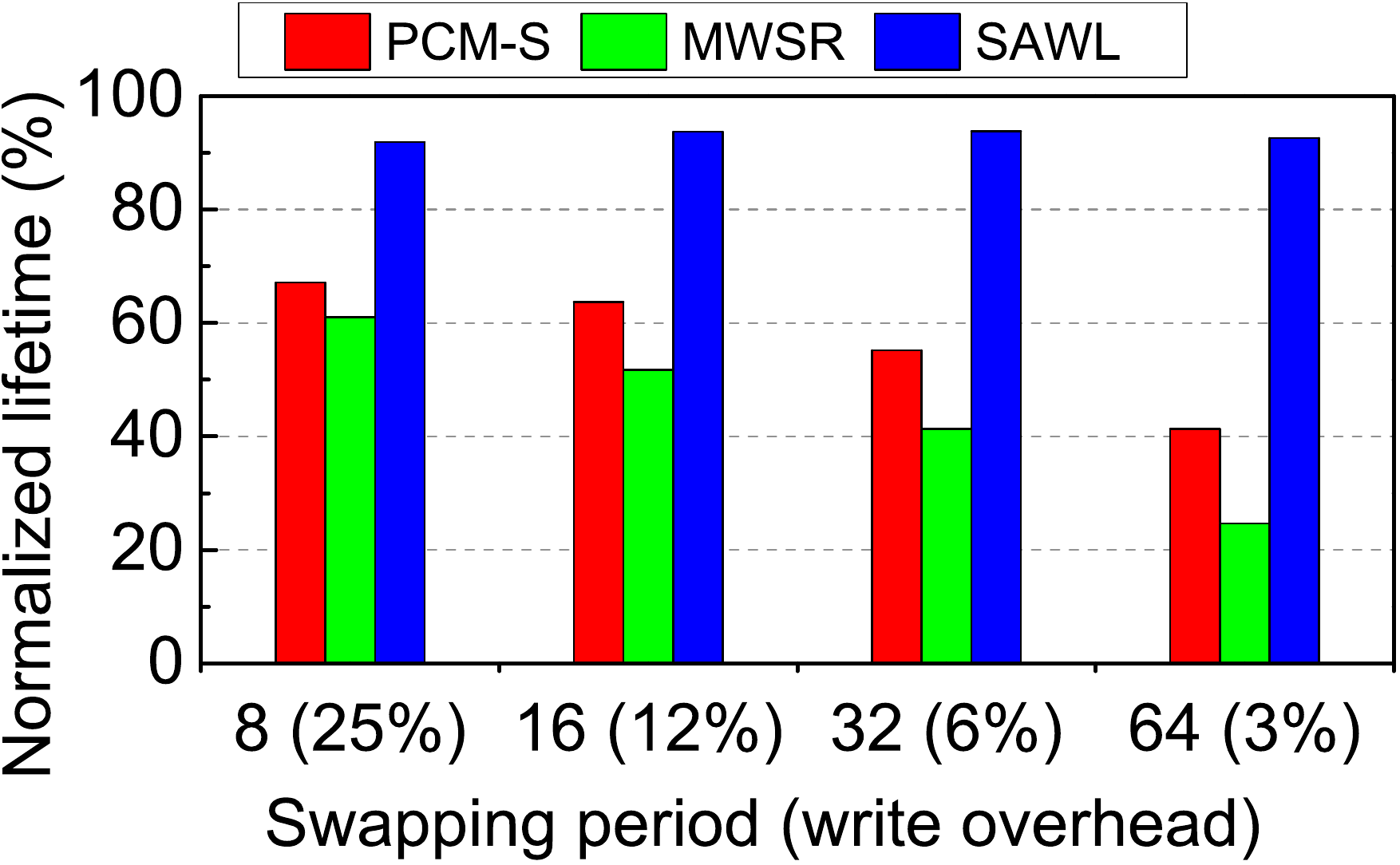}
    }
    \subfloat[$10^5$ endurance]{
    \label{fig:direct-parallel-c}
    \includegraphics [width=0.225\textwidth]{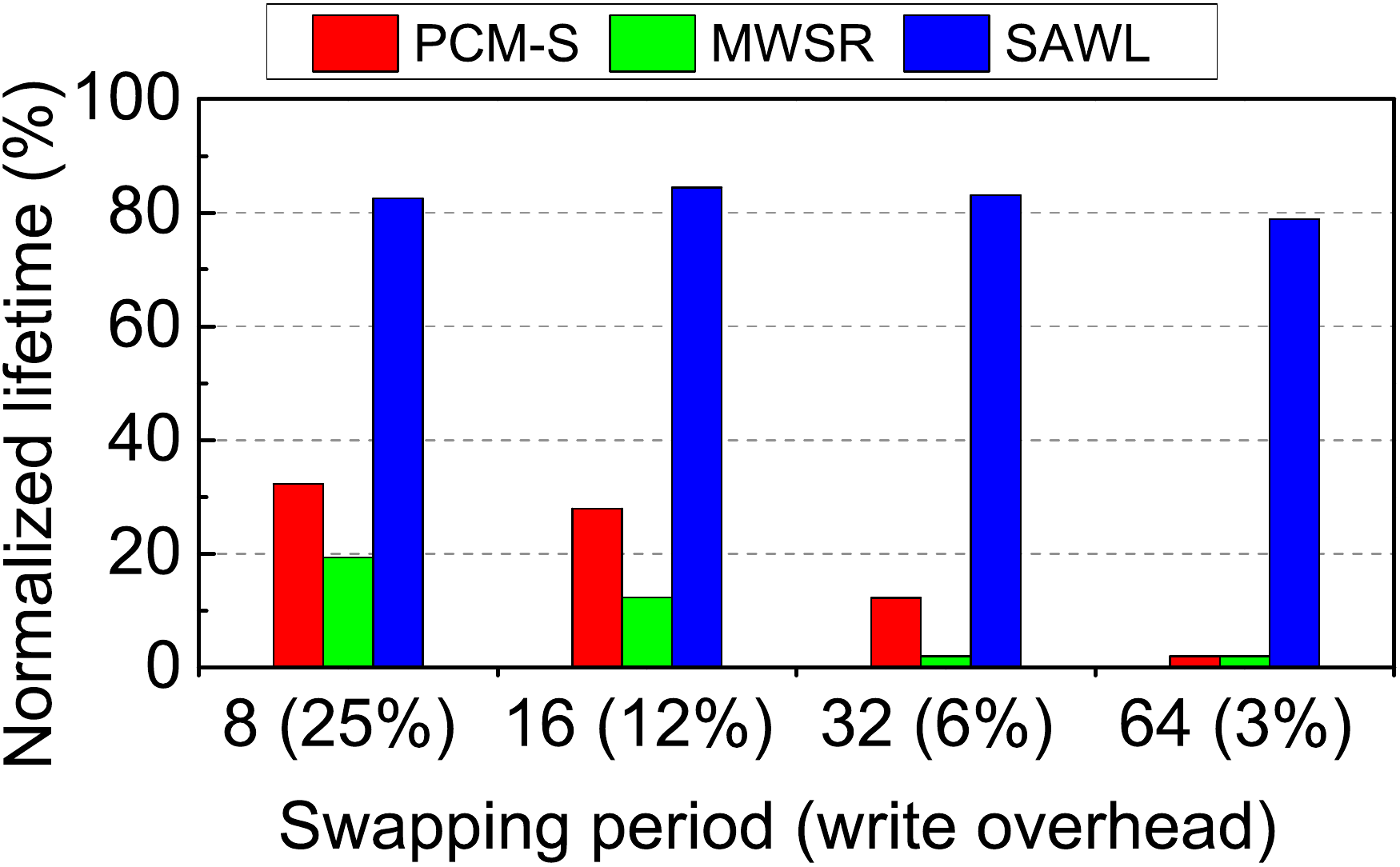}
    }
    \caption{\label{sawlmaliciouslifetime} The normalized lifetime of the MLC-based NVM system with PCM-S, MWSR and SAWL under different swapping periods.}
    \vspace{-0.5cm}
\end{figure}

{\textbf{1) NVM lifetime under the BPA program.} We use the BPA program to simulate the lifetime of an NVM system under the worst-case scenario and use the result to evaluate the robustness of the NVM system. We use 1MB on-chip cache and vary the swapping period from 8 to 64. The NVM lifetimes that PCM-S, MWSR, and SAWL are shown in Fig.~\ref{sawlmaliciouslifetime}. We observe that smaller swapping period increases the NVM lifetime for PCM-S and MWSR, but at the cost of high write overhead. SAWL achieves much higher lifetime than PCM-S and MWSR, due to storing all address mappings in NVM and no limitation on the number of regions. Fig.~\ref{sawlmaliciouslifetime} shows that SAWL improves $25\%\sim51\%$ ($50\%\sim78\%$) of ideal lifetime for the MLC-based NVM system with ${10}^{6}$ (${10}^{5}$) cell endurance, compared with PCM-S and MWSR.}

{\textbf{2) NVM lifetime under general applications.}} We evaluate the lifetime of an MLC-based NVM system under general applications. Since the requested address of the real-world workload changes every time, to evaluate the lifetime of an NVM system, the simulation must trace each request until the NVM system fails, which takes so much time that is unpractical. In order to reduce the running time, we simulate a 2GB NVM system with endurance of ${10}^{5}$. The normalized lifetime results can also be used to other large-capacity NVM systems. The entire space is divided into 4K $\sim$ 1M regions, and the exchange periods of TLSR, RBSG and SAWL algorithms are fixed at 128. Note that 4K regions are the standard configuration for TLSR and RBSG algorithms, and 1M regions are beneficial to our SAWL scheme.

\begin{figure}[t]  
\centering
\includegraphics[width=0.42\textwidth]{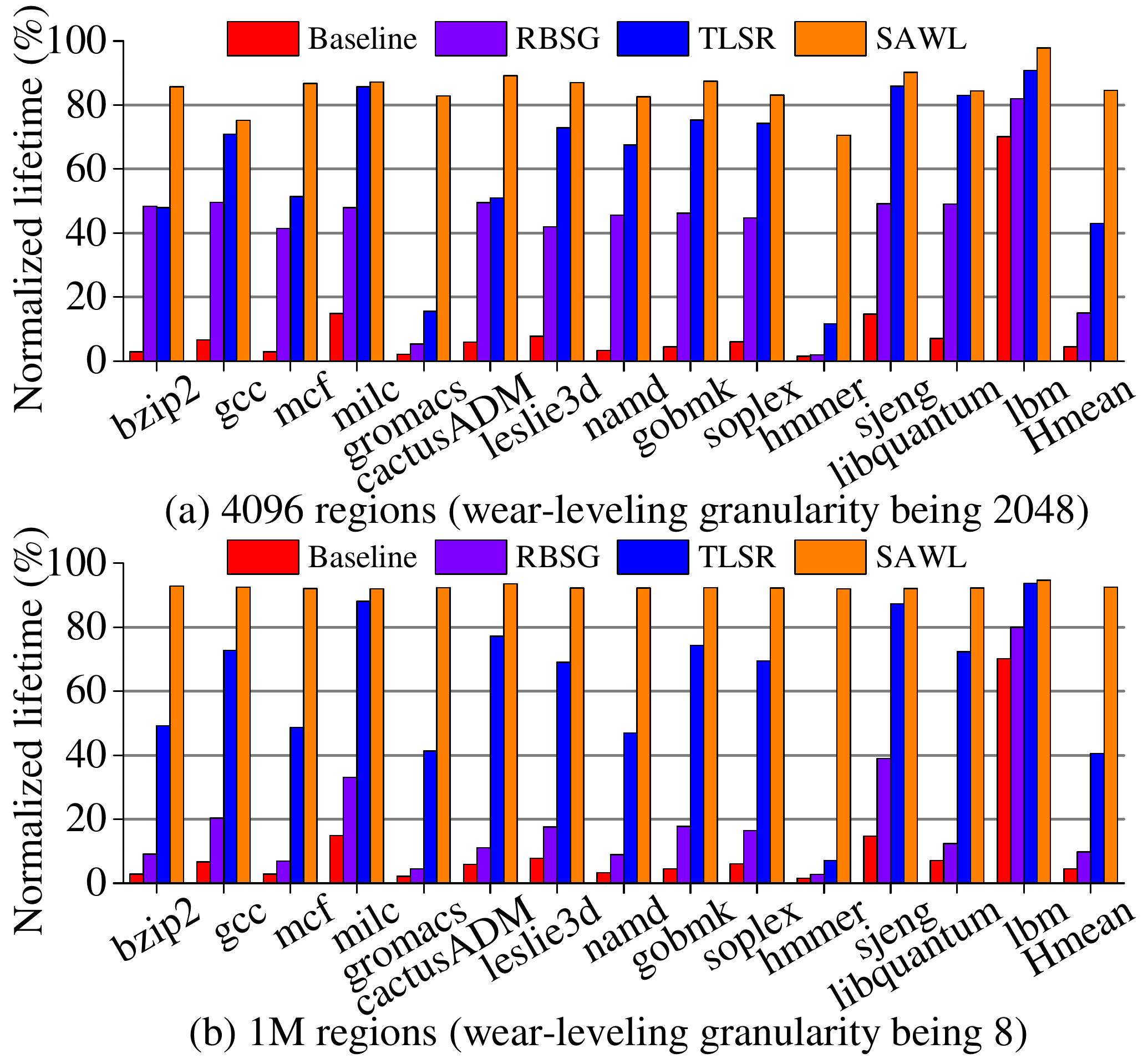}
\caption{The lifetime of the MLC-based NVM system with RBSG, TLSR and SAWL under general applications.}
\label{sawllifetime}
\vspace{-0.5cm}
\end{figure}

Fig.~\ref{sawllifetime} shows the normalized lifetime of the MLC-based NVM under the general benchmarks. The baseline system without any wear-leveling algorithm suffers from poor lifetime due to non-uniform underlying writes distribution. For the RBSG algorithm, the average lifetime (harmonic mean) of the MLC-based NVM system under all the benchmarks achieves 15\% of the ideal lifetime (ranges from 5\% to 81\%). The RBSG performs unsteadily since the static address mapping fails to balance inter-regional write distribution under various benchmarks. In contrast to RBSG, the results of the TLSR algorithm are much more stable for average lifetime, achieving an average lifetime that is 43.1\% of the ideal lifetime. What's worse, under $gromacs$ and $hmmer$ benchmarks, the lifetime of MLC-based NVM system decreases to 10\% of ideal lifetime, because the writes concentrate on a fraction of the address space. These experimental results clearly show that both the static and dynamical random address-mapping schemes are inadequate for most benchmarks. Compared to the existing wear-leveling algorithms, SAWL improves NVM lifetime to 85.1\% of the ideal lifetime. Under the most non-uniform distribution benchmarks, e.g., $gromacs$ and $hmmer$, SAWL still enhances NVM lifetime to 82\% and 70\%, respectively. In addition, the extra write overhead of the SAWL algorithm is less than 1\% and can be ignored. With the increase of the number of regions (1M regions), the SAWL algorithm can obtain higher lifetime, while the lifetime of RBSG and TLSRL is lower, as shown in Fig. \ref{sawllifetime} (b). The average lifetime of MLC-based NVM is extended to 9.8\%, 40.5\% and 92.5\% under RBSG, TLSR and SAWL schemes. In summary, the experimental results in Section 2.1 and 4.2 illustrate that the SAWL algorithm significantly improves the lifetime of the MLC-based NVM system under both malicious attacks and general applications.

\subsection{Performance Impact}
In general, a fine-grained wear-leveling region could improve lifetime but degrade performance. We compare NWL-4 (4-memory-line wear-leveling granularity on PCM-S and MWSR) with SAWL on cache hit rate and also compare NWL-4 and BWL with SAWL on IPC performance.

\begin{figure}[t]
\centering
\includegraphics[width=0.40\textwidth]{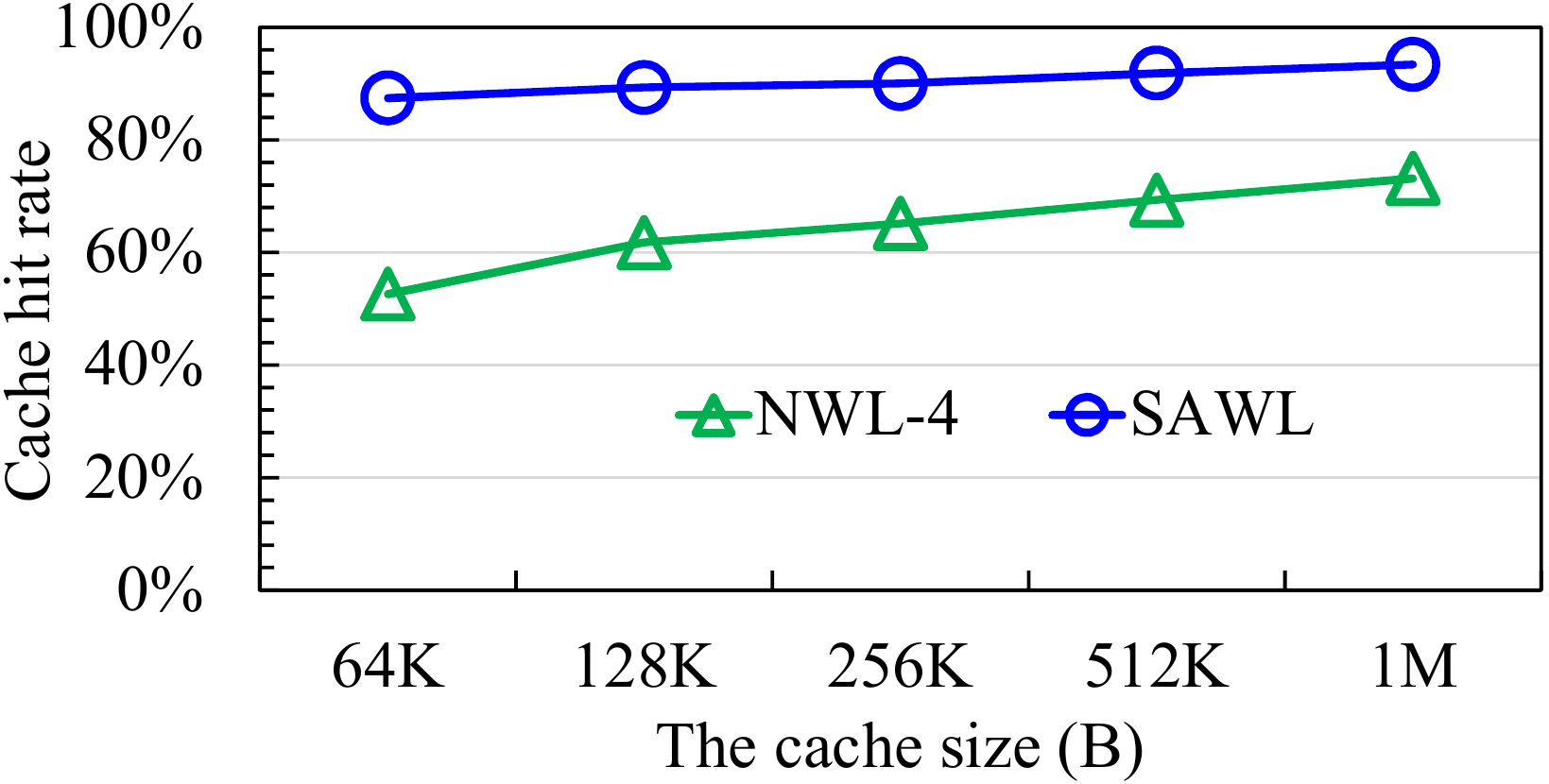}
\caption{The average cache hit rates of NWL-4 and SAWL with different cache sizes.}
\label{chr-nwl}
\vspace{-0.5cm}
\end{figure}

\textbf{1) Cache hit rate.}
To demonstrate the robustness of SAWL with different cache sizes, we evaluate the cache hit rate of SAWL as a function of the cache size ranging from 64KB to 1MB, compared with NWL-4. Fig. \ref{chr-nwl} shows {the average cache hit rates of the 14 applications from SPEC CPU2006. We observe that with the cache size increased from 64KB to 1MB, the average cache hit rate of NML-4 increases from 52\% to 73\%, and that of SAWL increases from 88\% to 94\%. Therefore, SAWL is able to achieve  $21\%\sim36\%$ of cache hit rate improvement via adaptively tuning the region sizes.}

\begin{figure}[t]
\centering
\includegraphics[width=0.38\textwidth]{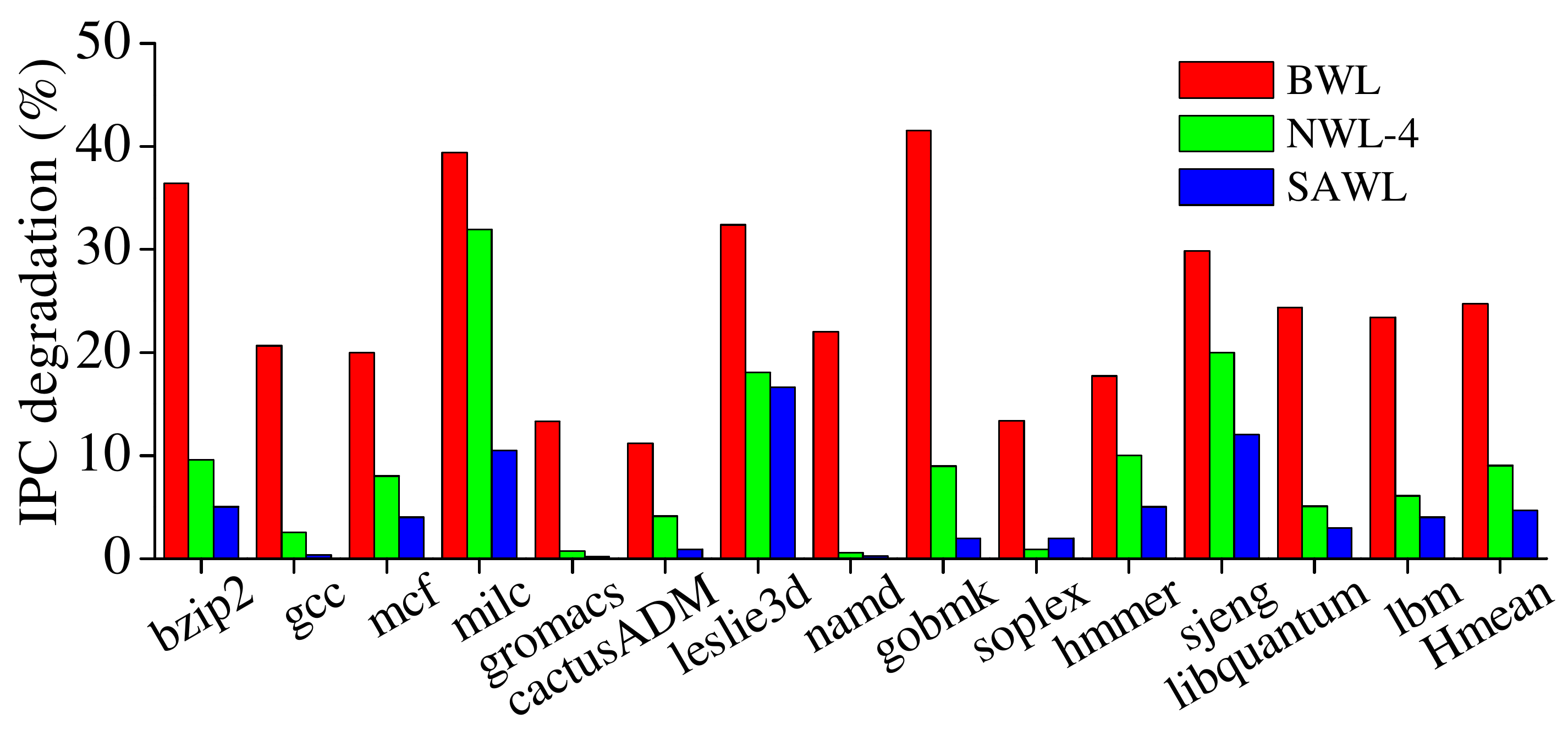}
\caption{IPC degradation of the NVM system (normalized to the baseline without wear leveling) with various wear-leveling schemes under the SPEC CPU2006 applications.}
\label{IPC-evaluation}
\vspace{-0.5cm}
\end{figure}

\textbf{2) IPC performance.}
We used the Gem5 simulator \cite{binkert2011gem5} to evaluate the performance impact of SAWL. In our experimental platform, the system consists of an 8-core processor (3.2 GHz), private 32KB L1 cache, shared 256 KB L2 cache, and a 128 MB L3 DRAM cache. The read and write latencies of DRAM are both 50ns, while that of MLC-based NVM (e.g., RRAM) are 50ns and 350ns, respectively. We use a queue length of 128 and the FR-FCFS scheduling scheme in the memory controller. The address translation requires 5 ns when the address is hit in the cache. Otherwise, it consumes 55ns. We run the 14 SPEC2006 applications mentioned above and compare the IPC measure with, i.e., normalized to, the Baseline (without any wear-leveling scheme). The swapping period of the SAWL algorithm is set to 128. As shown in Fig. \ref{IPC-evaluation}, the average IPC measure of the BWL, NWL-4 and SAWL schemes is decreased by 23\%, 10\% and 5\%, respectively. Some applications, such as the $bzip2$ and $milc$, show only slight IPC degradation. This is because the memory accesses in these applications are relatively sparse and the most requested addresses can be hit in the cache. Therefore, the results demonstrate that the performance impact of SAWL is arguably negligible.

\vspace{-0.3cm}

\section{Related Work}
Existing work relevant to our SAWL research can be broadly summarized into the two categories of wear leveling algorithms and writes reduction technologies.

\textbf{Wear Leveling Algorithms.}
Wear-leveling techniques attempt to balance write counts on physical devices. The conventional table-based wear-leveling algorithms include Fine-Grained Wear Leveling (FGWL)\cite{qureshi2009scalable}, row shifting and segment swapping\cite{zhou2009durable}, page allocation and page swapping \cite{chen2012age,ferreira2010increasing}, and line swapping \cite{yun2012bloom,gal2005algorithms}. To achieve long lifetime, the granularity of the mapping unit should be sufficiently small, which incurs huge space overhead. In addition, most of these algorithms adopt a deterministic exchanging policy, suffering from severe security vulnerability for malicious attacks. To address these problem, the algebraic-based wear-leveling algorithms, such as randomized region-based Start-Gap (RBSG) \cite{qureshi2009enhancing}, multi-level Security Refresh (TLSR) \cite{seong2010security} and Online Attack Detection (OAD) \cite{qureshi2011practical}, are proposed. RBSG and TLSR overcome the space overhead problem of TBWL algorithms, but the lifetimes under them are shortened for MLC-based NVM systems due to their insufficient data exchanges. OAD is used to tune the swapping frequency of AWL algorithms to improve NVM lifetime by distinguishing general applications with malicious attacks. Different from the pure algebraic algorithms, SAWL prolongs the NVM system to the ideal lifetime by tuning the wear-leveling granularities.

In addition, there are two hybrid wear-leveling algorithms combining table- and algebraic-based wear leveling, PCM-S \cite{seznec2010phase} and MWSR \cite{yu2012increasing}. PCM-S gathers multiple lines into a region and tracks the mapping information without recording the access frequencies of each region. During a write to a region, the region is swapped with a randomly picked region in memory with a small probability. A random amount of lines within the region are rotated during region exchange. However, the PCM-S and MWSR algorithms require a large number of regions to achieve uniform write distribution. The basic architecture that stores the entire address mapping table on the NVM devices leads to severe performance degradation. To overcome these problems, our SAWL leverages an SRAM cache in the memory controller and improve cache hit rate in runtime by tuning the region size dynamically, thus significantly alleviating performance degradation.

\textbf{Write Reduction Technologies.}
Write reduction technique is an architectural method to improve NVM lifetime. A hybrid design that consists of NVM-based main memory and a small-sized DRAM buffer is widely studied
\cite{dhiman2009pdram,qureshi2009scalable,ramos2011page,hu2013software,hu2015low,lee2014unified}. The DRAM buffers frequently re-write data and reduce the write traffic of NVM. Moreover, some techniques extend the lifetime via reducing the bit flipping, including Flip-N-Write redundant bit-write removal \cite{cho2009flip}, partial writes \cite{lee2009architecting}, line level write-back \cite{qureshi2009scalable}, lazy write \cite{qureshi2009scalable} and silent store removal \cite{zhou2009durable}. These techniques remove redundant writes by adopting read-before-write and novel coding schemes to extend NVM lifetime, which are orthogonal to our proposed wear-leveling technique.

\vspace{-0.3cm}

\section{Conclusion}
MLC technique can be used in the NVM systems, which leads to their rapid growth in device capacity but at the cost of much weaker endurance than their single-level-cell versions. The existing wear-leveling algorithms are shown to have their respective shortcomings for MLC-based NVM systems. While hybrid wear leveling has the potential to improve NVM lifetime, it incurs huge {on-chip} space overhead. The basic architecture, which stores the entire address mapping table on the NVM devices, leads to unacceptably severe performance degradation due to the very long address translation latency. To thoroughly address this problem, we propose a tiered wear-level architecture and a self-adaptive wear-leveling (SAWL) algorithm that dynamically tunes the wear-leveling granularities to accommodate more useful addresses in the cache, thus improving cache hit rate and system performance. Experimental results demonstrate that SAWL is effective and robust.
\vspace{-0.3cm}
{\normalsize \bibliographystyle{acm}
\bibliography{references}}


\end{document}